\documentclass[lettersize,journal]{IEEEtran}
\usepackage{amsmath,amsfonts}

\usepackage{algorithmic}
\usepackage{algorithm}
\usepackage{array}
\usepackage[caption=false,font=normalsize,labelfont=sf,textfont=sf]{subfig}
\usepackage{textcomp}
\usepackage{stfloats}
\usepackage{url}
\usepackage{verbatim}
\usepackage{graphicx}
\usepackage{xcolor}
\usepackage{csquotes}
\usepackage{makecell}
\usepackage[numbers]{natbib} 

\usepackage{tikz}
\usetikzlibrary{positioning,shapes.geometric}
\usepackage{adjustbox}
\usepackage{listings}
\usepackage{caption}

\newcommand{\xtheta}{x_{\theta}} 
\newcommand{\xzero}{x_{0}}       

\providecommand{\gls}[1]{#1}
\definecolor{rank1}{HTML}{70FF70}
\definecolor{rank2}{HTML}{858585}
\definecolor{rank3}{HTML}{454545}
\definecolor{rank4}{HTML}{000000}
\newcommand{\SIMSESpec}{\texttt{SIMSE\_Spec}}
\newcommand{\LoneSpec}{\texttt{L1\_Spec}}
\newcommand{\JTFS}{\texttt{JTFS}}
\newcommand{\DTWEnv}{\texttt{DTW\_Envelope}}
\newcommand{\LossSelect}{\textbf{Loss Selection}}
\newcommand{\SynthSelect}{\textbf{Synthesis Selection}}
\newcommand{\PeriodicLoss}{\textbf{Loss Landscape Navigation}}
\newcommand{\OutDomain}{\textbf{Out-of-Domain Generation}}
\newcommand{\BPNoise}{\textbf{BP-Noise}}  
\newcommand{\AddSineSaw}{\textbf{Add-SineSaw}}  
\newcommand{\AmpMod}{\textbf{Noise-AM}}  
\newcommand{\FMMod}{\textbf{SineSaw-AM}}  

\usepackage{xcolor}
\usepackage{tikz}

\lstdefinelanguage{Faust}{
    morekeywords={import, process, environment, declare, with, if, else, while, for, int, float, true, false},
    sensitive=true,
    morecomment=[l]{//}, 
    morecomment=[s]{/*}{*/}, 
    morestring=[b]", 
}

\begin{document}
\bibliographystyle{IEEEtran}

\title{Evaluating Sound Similarity Metrics for Differentiable, Iterative Sound-Matching}

\author{Amir Salimi, Abram Hindle, Osmar R. Za{\"i}ane}
\maketitle

\markboth{%
  \parbox{\textwidth}{%
    This work has been submitted to the IEEE for possible publication. \\
    Copyright may be transferred without notice, after which this version may no longer be accessible.%
  }%
}{}
 
\begin{abstract}
Manual sound design with a synthesizer is inherently iterative: an artist compares the synthesized output to a mental target, adjusts parameters, and repeats until satisfied. Iterative sound-matching automates this workflow by continually programming a synthesizer under the guidance of a loss function (or similarity measure) towards a target sound. Prior comparisons of loss functions have typically favored one metric over another, but only within narrow settings: limited synthesis methods, few loss types, often without blind listening tests. This leaves open the question of whether a universally optimal loss exists, or the choice of loss remains a creative decision conditioned on the synthesis method and the sound designer's preference. We propose differentiable iterative sound-matching as the natural extension of the available literature, since it combines the manual approach to sound design with modern advances in machine learning. To analyze the variability of loss function performance across synthesizers, we implemented a mix of four novel and established differentiable loss functions, and paired them with differentiable subtractive, additive, and AM synthesizers. For each of the sixteen synthesizer–loss combinations, we ran 300 randomized sound-matching trials. Performance was measured using parameter differences, spectrogram-distance metrics, and manually assigned listening scores. We observed a moderate level of consistency among the three performance measures. Our post-hoc analysis shows that the loss function performance is highly dependent on the synthesizer. These findings underscore the value of expanding the scope of sound-matching experiments, and developing new similarity metrics tailored to specific synthesis techniques, rather than pursuing one-size-fits-all solutions.
\end{abstract}

\section{Introduction}


A digital audio synthesizer is any software used for the creation and manipulation of audio. The possibilities afforded by these synthesizers are infinite, leading to their widespread adoption by artists and sound-designers~\cite{lyons1997understanding,russ1999sound,stranneby2004digital}. A typical synthesizer has a number of parameterizable functions which affect the sound output in various ways, and manual sound-design often involves modifying the parameters until a conceptualized target sound is reached. The automation of this approach has commonly been referred to as "sound-matching", with reduction of tinkering time and creation of ``interesting" sounds as the main motivators ~\cite{krekovic2019insights,turian2020sorry,horner1993machine,salimi2020make,esling2019flow,engel2020ddsp,mitchell2007evolutionary,shier2020spiegelib,masuda2021soundmatch,masuda2023improving}. The main requirements of sound-matching are a target sound, a similarity metric (or loss function), and a heuristic to find parameters for a synthesizer that replicate \textit{all} or \textit{some} of the characteristics of the target sound as best as possible~\cite{horner1993machine,mitchell2007evolutionary,masuda2023improving}. However, despite the seeming simplicity of the problem and decades of research, sound-matching is not yet a practical solution for sound-designers. Here we are motivated to understand why this is the case, and what future research should prioritize in order to improve the field. 

We provide a review of past works and identify major perennial issues in the field. In particular, we are concerned with the issues of \LossSelect~and \SynthSelect. The former refers to the search for an optimal loss function, lack of novel algorithms, and the---perhaps needless---focus on outperforming state of the art (\gls{SOTA}), while the latter refers to the lack of diversity in synthesis methods. As a consequence of these problems, we find that not much is known about the interaction between different methods of synthesis and different loss functions. Is there a universally best performing loss function for audio? Is there a need for further development of bespoke loss functions?

The main hypothesis here is that \textit{the performance of a similarity measure (or loss function) is influenced by other factors in the environment, particularly the method of synthesis}. Testing this hypothesis requires a variety of sound-matching experiments that measure whether a single loss function proves most effective. While there have been many works and experiments comparing the accuracy of loss functions~\cite{vahidi2023mesostructures,turian2020sorry,engel2020ddsp,uzrad2024diffmoog,han2023perceptual,masuda2021soundmatch,turian2020sorry,bruford2024synthesizer}, we note that claims regarding the effectiveness of one function versus another have often been made in limited contexts that may not generalize to other settings. Beyond this central hypothesis, we also investigate three additional research questions: (Q1) To what extent do automatic evaluation metrics agree with manual listening tests? (Q2) Can loss functions based on Dynamic Time Warping (DTW) and Scale-Invariant Mean Squared Error (SIMSE) provide advantages over SOTA loss functions, and under what conditions?
(Q3) Is iterative differentiable optimization a viable strategy for design of sound-matching experiments?

 We adopt a lesser used \textit{iterative} and \textit{differentiable} approach to defining sound-matching experiments. The iterative approach better mimics the manual process of recursive listening and parameter adjustments towards sound design, while a differentiable environment allows access to the loss function gradients that can be used to better understand the nature of the problem. We define four differentiable synthesizers (each showcasing a fundamental method of synthesis used in modern synthesizers) and pair them with four different loss functions (two established methods, one utilizing DTW, and one utilizing SIMSE). We evaluate the final similarity with two different automatic methods, as well as manual listening tests conducted by two of the authors. We apply statistical ranking to the evaluation scores to compare outcomes across synthesizer–loss combinations and measure the similarity of different evaluation measures. 

\textbf{Contributions.} The contributions of this paper include (1) an evaluation of multiple differentiable losses across multiple synthesis methods, (2) the introduction and justification of loss functions utilizing DTW and SIMSE (3) a discussion on the utility and agreement between manual and automatic evaluation metrics (4) a nomenclature of different sound-matching approaches and unsolved issues in the field.



\section{Background and Related Work}
\label{sec:background_related}
 Sound-matching sits at the intersection of digital signal processing, audio representation, and optimization. Here we first formalize the sound-matching task, then review core synthesis methods, similarity measures, and optimization approaches. 
We conclude with a survey of related work and a discussion of gaps in the field. 

\subsection{Formalization of Sound-Matching}
\label{sec:sound_matching_definition}
Following prior works~\cite{vahidi2023mesostructures,han2023perceptual}, we define sound-matching in terms of a parametric synthesizer, a target sound, a representation function, and a similarity measure. 
The key components are:
\begin{itemize}
    \item $g(\theta)$: A parametric audio synthesizer $g$ with parameters $\theta$.
    \item $x_\theta$: The synthesizer output, $x_\theta = g(\theta)$.”.
    \item $\xzero$: The target sound to be replicated or imitated. 
    \item $\phi(\cdot)$: A representation (feature extraction) function mapping signals into a comparison space.
    \item $L$: A loss function that measures distance between $\xtheta$ and $\xzero$, typically via $L(\theta,\xzero) = d(\phi(\xtheta), \phi(\xzero))$ for some metric $d$.
\end{itemize}

This formalization highlights that sound-matching is not a single well-defined optimization problem, but rather depends on design choices for $g$, $\phi$, and $L$. 
Moreover, the ``correct'' solution may not exist in a strict sense: depending on the artistic goal, multiple parameter settings may yield acceptable or even preferable outputs.  In the following sections we will discuss the various components of sound-matching, the subjectivity of ``correct'' solutions, and important issues in the field.

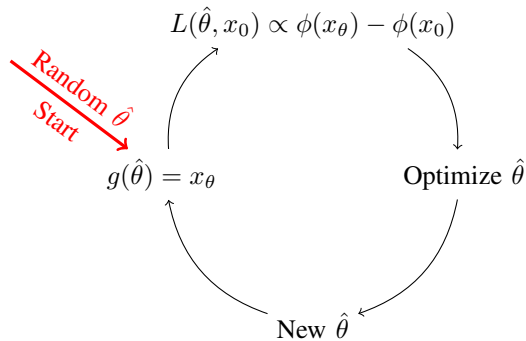
\begin{figure}[ht]
    \centering
\begin{tikzpicture}[node distance=2cm, auto]

\node (start) [text centered] {\( g(\hat{\theta}) = x_{\theta} \)};
\node (L) [above of=start, right of=start, text centered] 
    {\( L(\hat{\theta}, x_{0}) \propto \phi(x_{\theta}) - \phi(x_{0}) \)};
\node (optimize) [below of=L, right of=L, text centered] {Optimize $\hat{\theta}$};
\node (new_theta) [below of=optimize, right of=start, text centered] {New \( \hat{\theta} \)};

\draw[->, very thick, red] (start) ++(-2,1.5) -- (start)
    node[midway, below, align=center, sloped, color=red] {Start}
    node[midway, above, align=center, sloped, color=red] {Random $\hat{\theta}$};

\draw[->, bend left] (start) to node[midway, right, align=center] {} (L);
\draw[->, bend left] (L) to node[midway, right, align=center] {} (optimize);
\draw[->, bend left] (optimize) to node[midway, below, align=center] {} (new_theta);
\draw[->, bend left] (new_theta) to node[midway, left, align=center] {} (start);

\end{tikzpicture}
    \caption{ Iterative approach to sound design.}
    \label{fig:sound_design_loop_iterative}
\end{figure}

 Figure~\ref{fig:sound_design_loop_iterative} shows a general model for iterative sound-matching. To begin the process, the parameters are arbitrarily initialized (usually random generation), the similarity of the target and output is measured, and the parameter is \textit{optimized} with the goal of increasing the similarity, or alternatively, reducing the loss. This process repeats until termination. 

\subsection{Digital Signal Processing and Synthesis}
\label{sec:dsp}

A digital audio synthesizer generates or processes audio by chaining digital signal processing (\gls{DSP}) functions. 
Each function is parameterized, and the set of parameters defining a chain of DSP functions is called a synthesizer program. 

The simplest DSP function is a sinusoidal oscillator: 
\[
x[n] = \sin\!\left(2 \pi f \frac{n}{SR}\right),
\]
where $f$ is frequency in hertz, $SR$ is the sampling rate, and $n$ is the discrete time index. 
At a sampling rate of $SR$ samples per second, a 1\,Hz oscillator completes one cycle per second. 
Such computations form the foundation of digital audio synthesis~\cite{lyons1997understanding}. 

Since the advent of DSP in the 1960s~\cite{stranneby2004digital}, a wide variety of parametric functions have been proposed, including oscillators, filters, equalizers, and envelopes~\cite{lyons1997understanding,russ1999sound}. 
Sound design involves modifying these parameters until a desired output is reached~\cite{roads1996computer,pinch2004analog}. 

Early sound-matching research focused heavily on frequency and amplitude modulation (\gls{FM}/\gls{AM}) synthesis~\cite{horner1993machine,mitchell2007evolutionary,vahidi2023mesostructures}, which is simple to implement yet expressive~\cite{chowning1973synthesis}. 
Other synthesis methods studied in \textit{isolation} include additive and subtractive synthesis~\cite{engel2020ddsp,masuda2023improving,salimi2020make} and physical modeling~\cite{riionheimo2003parameter,han2024learning}. By \textit{isolation}, we mean settings where the effect of individual parameters on the output sound remains tractable. For example, we exclude studies using commercial Virtual Studio Technology (VST) which can obscure the interactions between modules, losses, and outputs. 

Finally, recent years have seen a growing interest in differentiable DSP (\gls{DDSP})~\cite{engel2020ddsp}, which integrates gradient-based optimization~\cite{goodfellow2016deep,boyd2004convex} with DSP building blocks. 
Implementing complex DSP functions in a differentiable manner remains challenging, and robust differentiable audio similarity measures are still under active investigation~\cite{masuda2021soundmatch,vahidi2023mesostructures,uzrad2024diffmoog}.

\subsection{Sound Representation and Loss Functions}
\label{sec:loss_funcs}
A digital sound (or an audio signal) is a series of numbers~\cite{smith1991viewpoints,smith2007mathematics}. To compare two digital sounds, the two corresponding series are passed to a function that measures their similarity. Two signals can sound identical to our ears, without having any values in common~\cite{moore2012introduction}. This necessitates the use of proxy representations (or feature extractors) when comparing sounds automatically. Similarity between the target sound and the synthesizer output is then measured by some form of subtraction and summation of the proxy representations.

In sound-matching, particularly in a Deep Learning (\gls{DL}) context~\cite{goodfellow2016deep}, the similarity function can also be called a \textit{loss} function, where the emphasis is on measurement and reduction of the distance between target and output. It is important to note that there is a close relationship between the loss function $L$ and the sound representation function $\phi$. $L$ is the result of a distance measure $d$ applied to the features extracted by $\phi$. 

\[
L(\theta, \xzero) = d(\phi(\xtheta), \phi(\xzero))
\]

\noindent

A proxy representation is the output of the function \( \phi \), which can be thought of as a feature extraction function that maps the sounds  $\xzero$ and $\xtheta$  to their respective representations. 
The proportionality or distance metric $d$ has typically been the L1 or L2 distance~\cite{turian2020sorry,richard2025model}, with L1 being calculated as the mean of the absolute difference between every point in the proxy representation:
\[
L(\theta, \xzero) = \left\| \phi(\xzero) - \phi(\xtheta) \right\|_1
\]

Here we discuss four methods of audio representation and the corresponding loss functions, including technical justifications for our novel methods.

\subsubsection{Parameter Loss}
A common measure of similarity in sound-matching is the distance between synthesizer parameter sets, referred to as ``P-Loss"~\cite{han2023perceptual}. Typically, for the implementation of P-Loss the parameter sets are treated as vectors in space, and L1 or L2 distance is applied. There are two major limitations to this approach: First, the target and output sound must be made by the same synthesizer; otherwise the parameter sets cannot be compared (see Section~\ref{sec:in-domain}). Second, the relationship between synthesizer parameters and the audio output is not linear~\cite{shier2020spiegelib,han2023perceptual,esling2019flow}. 

\subsubsection{Fourier Spectrograms}
\label{sec:fourier_specs}
Fourier-based transformations such as short-time Fourier transforms (\gls{STFT}), Mel-spectrograms, and Mel-frequency cepstral coefficients have been viewed as the de facto and state-of-the-art representation of audio~\cite{beauchamp2003error,mitchell2007evolutionary,yee2018automatic}, however, there are many issues associated with their use in sound-matching~\cite{turian2020sorry,vahidi2023mesostructures,han2023perceptual,uzrad2024diffmoog}. Fourier transformations allow for the conversion of a signal from the time-domain to the frequency domain. Audio spectrograms can be generated by segmentation of a piece of audio into overlapping windows followed by the application of Fourier transforms to each window. They are costly to compute, but provide a better temporal view of changes in frequency content~\cite{muller2007dynamic,smith2007mathematics}. There are different types of spectrograms that have a basis in Fourier transformations, but the most notable and commonly used is the STFT.  
What we call \textit{Fourier-based Spectrograms} are variations on the STFT approach. For example, Mel-Spectrograms \textit{bin} frequencies on a near-logarithmic scale to better match human perception of frequencies~\cite{muller2007dynamic}. Multi-scale spectrograms (\gls{MSS}) used in recent works are a simple weighted average of multiple spectrograms with different parameters such as window size, number of frequency bins, and hop size~\cite{engel2020ddsp,vahidi2023mesostructures}; this may provide some improvements at a higher computational cost~\cite{turian2020sorry,engel2020ddsp}.

A common limitation of spectrogram-based losses is their sensitivity to global gain: two signals with identical spectral shape but different overall amplitude can yield large errors under L1/L2 norms. 
Here we propose and test Scale-Invariant Mean Squared Error (SIMSE) as an alternative to L1/L2 norms in spectrogram comparisons. SIMSE normalizes the amplitudes before comparison and emphasizes proportional differences in spectral envelopes~\cite{barron2014shapessimse}. 
Perceptually, listeners are often tolerant of loudness changes while remaining sensitive to timbral shape, therefore this property could be advantageous for subtractive synthesis, where filter cutoffs reshape spectral balance without predictable changes in total energy. Although SIMSE has been applied in other contexts such as audio reconstruction, to our knowledge its use as a differentiable spectrogram loss in iterative sound-matching is novel.


\subsubsection{Joint-Time Frequency Spectrum}
Recent works have focused on the limitations of parameter and spectral loss functions in sound-matching~\cite{vahidi2023mesostructures,uzrad2024diffmoog}, seeking to create more effective general solutions for the comparison of audio. 
Noting the weaknesses of comparing STFT spectrograms (such as alignment and loudness distances), Vahidi \textit{et al.} proposed differentiable Joint-Time Frequency Scattering (\gls{JTFS})~\cite{anden2015joint} as an alternative to spectrogram loss in sound-matching, and showed improved performance in sound-matching with differentiable chirplet synthesizers~\cite{vahidi2023mesostructures}. JTFS is the result of the application of a 2D wavelet transformation to the time-frequency representation of a signal~\cite{anden2015joint} and has been reported as more sensitive to \textit{mesostructural} features such as melody, syncopation, and textural
contrast~\cite{vahidi2023mesostructures}.

\subsubsection{Dynamic Envelope Warping}
Dynamic Time Warping (DTW) is a method for measuring similarity between multi-dimensional time-series~\cite{rabiner1993fundamentals,muller2007dynamic,giorgino2009computing}. Given any two time-series $X = \{x_1,x_2,...,x_m\}$ and $Y = \{y_1,y_2,...,y_n\}$, we have indices $i\in\{1...m\}$ and $j\in\{1...n\}$ defining $X$ and $Y$. When the series are \textit{warped}, these indices change to expand or contract different portions of the series. To borrow the notation given by Muller~\cite{muller2007dynamic}, warped indices are a sequence $p=(p_1,...,p_L)$, where \(p_\ell = (m_\ell, n_\ell) \in [1 : m] \times [1 : n] \text{ for } \ell \in [1 : L]\), meaning that the indices for $X$ and $Y$ are reorganized under special conditions. In classical DTW, these conditions are \textit{monotonicity}, \textit{boundary matching}, and \textit{single step-size}. DTW measures the distance between the time-series \textit{after} alignments, typically using Euclidean distance, such that the distance between a time-series and shifted versions of itself would be 0, regardless of shift amount~\cite{tavenard.blog.dtw}. Additional rules can be imposed to keep alignments locally constrained~\cite{itakura1975minimum,sakoe1978dynamic}.

DTW provides robustness to local temporal shifts, making it well suited for comparing modulated signals where the perceptual similarity lies in envelope dynamics rather than precise onset alignment. 
For example, two tremolo signals with slightly different phase are perceptually similar but would appear distant under spectrogram L1. 
A possible use case to consider is the application of DTW to amplitude envelopes, which may be able to match two sounds with similar rates of loudness modulation regardless of alignments. 
To our knowledge, DTW applied in this way has not been used as a loss in sound-matching (whether differentiable and iterative or not).

\subsection{In-Domain Versus Out-of-Domain}
\label{sec:in-domain}
The choice of domain depends on whether we want to use the same synthesizer for the target and output sounds, the scenario that is called \textit{in-domain}, or have target sounds that came from sources other than the synthesizer, or \textit{out-of-domain}. To paraphrase the description given by Masuda \textit{et al.}~\cite{masuda2021soundmatch}, if $g$, the synthesizer of choice, can accurately replicate the target sound $\xzero$, or put differently, if $\xzero$ itself is an output of $g$, then the sound-matching task is \textit{in-domain}. If $\xzero$ is not an output of $g$, then the sound-matching task is \textit{out-of-domain}. In-domain tasks in general are simpler, and often there is a guarantee that there is a correct answer to the sound-matching problem, particularly if the goal is accurate replication of the sound. If the target sound is out-of-domain, replication is not guaranteed, and the goal becomes the \textit{imitation} of some aspect of sound. 

Regardless of the domain, the generation goal can be \textit{replication} or \textit{imitation} of the target sound. In replication, the goal is to make an identical copy of the target sound. Imitation is an artistic pursuit and harder to define, since the goal is to make new sounds that only retain a subset of the target's sonic features. While closely related to the in-domain versus out-of-domain problem, the choice of replication versus imitation is more dependent on how the loss and representation functions are defined. 

\subsection{Supervised Versus Direct Optimization}
\label{sec:optimization}

The choice of \textit{heuristics} is yet another important attribute sound-matching. The goal of sound-matching is to find the optimal parameters $\theta^*$ that minimize the loss between the synthesizer output and the target sound. 
\[
\theta^* = \arg\min_{\theta} L(\theta,\xzero)
\]

The heuristics (i.e., how $\theta^*$ is approximated) used in past works can be broadly split into the two categories of \textit{direct optimization} and \textit{supervised}  (or inference) methods. Direct optimization refers to the iterative generation of a sound output, measurement of similarity between target and output, and application of updates to the parameters to maximize similarity (or minimize loss)~\cite{horner1993machine,mitchell2007evolutionary,yee2018automatic,vahidi2023mesostructures}; while supervised methods use large datasets of synthesizer sounds and corresponding parameters to learn the generation objective, commonly with the use of DNNs~\cite{engel2020ddsp,salimi2020make,yee2018automatic,esling2019flow}. These models often make their parameter predictions in a single step (or 1-shot). 

Genetic algorithms (\gls{GA})~\cite{holland1992genetic} have been the earliest and most common heuristic for direct sound-matching~\cite{horner1993machine,mitchell2007evolutionary,yee2018automatic}. These algorithms start with arbitrary parameter sets that can be treated as an evolving population where the genomes are the parameter values. The most fit members of the group are the parameters which perform the best in the loss function, and create a new generation of parameters via mutation (random change in subset of parameters) and crossovers (combination of parameter sets); this process repeats until the goal or a maximum number of generations is reached. Rather than using random mutations, differentiable approaches allow goal-oriented updates to the synthesizer parameters (the goal being the minimization of loss), but with some drawbacks: other than requiring more computation power, differentiable functions require careful implementation of operations that are continuous and numerically stable; this makes the implementation of signal processing functions quite difficult, contributing to the scarcity of works in this domain.

\begin{table*}[t]
\centering
\caption{Summary of select works in sound-matching. Gen. refers to whether the generations are done in 1-shot (generally, output by a neural network) or a search is conducted with the goal of iteratively getting closer to the target sound.}
\setlength{\tabcolsep}{3pt} 
\renewcommand{\arraystretch}{1.05} 
\scriptsize 
\resizebox{\textwidth}{!}{
\begin{tabular}{|>{\raggedright\arraybackslash}m{2cm}|>{\raggedright\arraybackslash}m{2cm}|>{\raggedright\arraybackslash}m{2cm}|>{\raggedright\arraybackslash}m{2cm}|>{\raggedright\arraybackslash}m{1.2cm}|>{\raggedright\arraybackslash}m{1.2cm}|>{\raggedright\arraybackslash}m{1cm}|>{\raggedright\arraybackslash}m{1cm}|}
\hline
\textbf{Work} & \textbf{Synthesis} & \textbf{Loss} & \textbf{Heuristics} & \textbf{Goal} & \textbf{Domain} & \textbf{Year} & \textbf{Gen. } \\
\hline
Horner \textit{\textit{et al.}}~\cite{horner1993machine} & Non-differentiable FM & STFT (McAulay-Quatieri) & GA & Replicate & In & 1991 & Iter \\
\hline
Mitchel \textit{\textit{et al.}}~\cite{mitchell2007evolutionary} & Non-differentiable FM & Spectral Relative & Evolutionary & Replicate & In (Contrived) & 2007 & Iter \\
\hline
Yee-King \textit{\textit{et al.}}~\cite{yee2018automatic} & Non-differentiable VST & P-Loss (for Supervised), Spectral (MFCC) & Supervised \& Evolutionary & Replicate & In & 2018 & Both \\
\hline
Esling \textit{\textit{et al.}}~\cite{esling2019flow} & Non-differentiable VST & Spec MSS/SC P-Loss & Supervised \& Modeling & Replicate & In* \& Out & 2019 & 1-shot \\
\hline
Shier \textit{\textit{et al.}}~\cite{shier2020spiegelib} & Custom, non-diff & FFT \& Spectral & Supervised \& Direct & Replicate & In & 2020 & Both \\
\hline
Salimi \textit{\textit{et al.}}~\cite{salimi2020make} & DSP & STFT and Envelope & Supervised & Imitate & Out & 2020 & 1-shot \\
\hline
Masuda \textit{\textit{et al.}}~\cite{masuda2021soundmatch} & DDSP & P-Loss initially, fine-tune with Spec. & Supervised & Both & In \& Out & 2021 & 1-shot \\
\hline
Han \textit{\textit{et al.}}~\cite{han2023perceptual} & NN-encoder $\rightarrow$ physical model & PNP (approx. of STFT) & Supervised & Replicate & In & 2023 & 1-shot \\
\hline
Vahidi \textit{\textit{et al.}}~\cite{vahidi2023mesostructures} & Differentiable FM & JTFS/MSS & Gradient desc. & Replicate & In & 2023 & Iter \\
\hline
Barkan \textit{et al.}~\cite{barkan2023inversynthII}& Diff. Synthesizer Proxy& P-Loss + L1 Spec (weighted) &  Supervised (init.), Semi-Supervised & Replicate& In & 2023 & Both
\\
\hline
Uzrad \textit{\textit{et al.}}~\cite{uzrad2024diffmoog} & DDSP & Synth. Chain and P-Loss & Supervised & Replicate & In \& Out & 2024 & 1-shot \\
\hline
Cherep \textit{et al.}~\cite{creativecherep2024} & DDSP & CLAP & Evolutionary & Imitate & Out & 2024 & Iter \\
\hline
\end{tabular}
}

\label{tab:summary}
\end{table*}

\subsection{Historical Framing of Sound-Matching}
Given the attributes discussed above, Table~\ref{tab:summary} summarizes relevant literature. Here we expand with a historical analysis of past sound-matching work.

Perhaps the earliest foundational study is Justice~\cite{justice1979analytic}, who analytically decomposed and recreated sounds using a simple FM synthesizer (see Section~\ref{sec:dsp}). 
Subsequent work retained this FM structure but explored new heuristics for parameter search. 
For example, Horner \textit{et al.}~\cite{horner1993machine} applied genetic algorithms (GAs) to resynthesize sounds with one modulator and one carrier, measuring similarity via the McAulay–Quatieri method~\cite{mcaulay1986speech}.

Later studies introduced more complex synthesis models, such as wavetables~\cite{horner2003auto} and physical modeling~\cite{riionheimo2003parameter}. 
Mitchell and Creasy~\cite{mitchell2007evolutionary} highlighted the difficulty of disentangling synthesizer limitations from optimization inefficiency. 
They proposed a \textit{contrived methodology} in which the best heuristic for in-domain FM resynthesis should also generalize to out-of-domain targets (e.g., muted trumpet tones~\cite{opolko1989mcgill}). 
However, limited testing produced contradictory results, and they concluded that changes to the synthesizer, loss, or sound domain effectively redefine the search space~\cite{mitchell2007evolutionary}.

Recent years have seen more works in sound-matching using supervised machine learning techniques. In 2018, Yee-King \textit{et al.} rendered 60,000 audio-parameter pairs from the \textit{Dexed} \gls{VST} synthesizer~\footnote{https://asb2m10.github.io/dexed/}, and showed that NNs trained on this dataset can outperform GA and hill-climber (\gls{HC})~\cite{hoffmann2000heuristic} methods in rendering speed, with slight improvements in MFCC error (used as an objective performance test). The speed improvement appears trivial, considering the iterative nature of GAs when compared to offline training of supervised models. For GAs and HC optimizer, MFCCs were used as a measure of performance as well as a loss function. For training the networks, P-Loss was used, since differentiable MFCCs were not possible in their pipeline. Importantly, informal listening tests found the results unsatisfactory, likely due to the synthesizer’s 155-parameter complexity. 

Masuda \textit{et al.} also applied supervised learning~\cite{masuda2021soundmatch}. Their work highlights the issue of non-linearity in parameter-to-synthesizer outputs and out-of-domain search. This work uses a differentiable subtractive synthesizer (two oscillators and an LP filter) alongside a NN model pre-trained with P-Loss on an in-domain dataset of randomly selected parameters. After training, the model was fine-tuned using 20,000 out-of-domain sounds from the NSynth dataset~\cite{engel2017neural} and multi-scale spectrogram loss~\cite{engel2020ddsp}. This approach proved more effective---i.e., achieved lower multi-level spectral difference in out-of-domain tests---than baseline models, which were either not exposed to out-of-domain sounds or trained exclusively with P-Loss. Subjective hearing tests were conducted, showing a preference for the fine-tuned model~\cite{masuda2021soundmatch}. Masuda \textit{et al.} later extended this work with semi-supervised learning, highlighting significant gaps in in-domain and out-of-domain performance~\cite{masuda2023improving}.

Rather than focusing on a particular implementation, Shier \textit{et al.} presented Spieglib, a library for implementation of sound-matching pipelines~\cite{shier2020spiegelib}. This library provides different choices for DNNs, GAs, synthesizers, and feature extractors. Shier \textit{et al.} presented an experiment with a similar setup to Yee-King \textit{et al.}~\cite{yee2018automatic}, however, they found a genetic algorithm as the best performing.

Differentiable loss functions that use spectrogram differences can be computationally expensive. To mitigate this, Han \textit{et al.}~\cite{han2023perceptual} introduced ``perceptual-neural-physical loss'' (PNP). PNP is an approximation of loss functions; specifically, a loss function that uses the L2 norm of the difference between the features of two sounds, or $||\phi(t) - \phi(x)||^2_2$, where $\phi$ could be a spectrogram or JTFS function. PNP loss functions are fast and differentiable, but require training and parameter estimation. A Riemannian metric M needs to be calculated for the minimization of locally linear approximation of the ``true" spectral loss function~\cite{han2023perceptual}. 

\[
\| \varphi(\xzero) - \varphi(\xtheta) \|_2^2
= \langle \tilde{\theta} - \theta \,|\, M(\theta) \,|\, \tilde{\theta} - \theta \rangle
+ O(\|\tilde{\theta} - \theta\|_3^2).
\]

This metric is calculated alongside the neural parameter estimator. Once trained, PNP enabled the fast differentiable optimization of computationally expensive loss functions such as JTFS with FM and physical models~\cite{han2023perceptual,han2024learning}.

A neural approximation for another part of the supervised sound-matching chain---this time the synthesizer---was proposed by Barkan \textit{et al.}~\cite{barkan2023inversynthII}. As they noted, without a differentiable synthesizer, ``model-based'' (or what we call supervised) approaches cannot directly compare $\xtheta$ to $\xzero$, often opting for P-Loss, which may not correctly map the parameters of sound to the output audio~\cite{esling2019flow,han2023perceptual,masuda2023improving}. This approach used a model for mapping sounds to parameters, another for approximating the synthesizer, and a loss function which combines P-Loss with STFT differences. This ``Inversynth II'' (IS2) approach yielded significant improvements to previous works which did not use the STFT approximations for loss~\cite{esling2019flow,barkan2019inversynth}. Barkan \textit{et al.} then attempted to improve the IS2 model with Inference-Time
Fine-tuning (ITF). For ITF, the synthesizer approximation is frozen and the encoder is iteratively fine-tuned for a particular sample. However, ITF often degraded performance, possibly due to proxy–synthesizer mismatch or overfitting.

Uzrad \textit{et al.}~\cite{uzrad2024diffmoog} took another unique approach to sound-matching: using a differentiable \textit{synthesis chain} of DSP generators and effects and a loss function that combines P-Loss with a \textit{signal-chain loss}. The synthesizer is a customizable chain of effects, which feed one output as input to the next step of the chain; signal-chain loss compares the parameter and output difference at every output step in the chain~\cite{uzrad2024diffmoog}. Possible chain functionalities are FM/AM, Low-Frequency Oscillators (\gls{LFO}), filters, and envelopes. Like the results shown by Masuda \textit{et al.}~\cite{masuda2021soundmatch}, better out-of-domain results were achieved when pre-trained on in-domain data and fine-tuned using out-of-domain NSynth data~\cite{engel2017neural}.

 Audio embeddings can also be used as a similarity metric. In a recent work, Cherep \textit{et al.} used latent representations from the CLAP model~\cite{wu2023large} along with a differentiable synthesizer~\cite{synthhaxcherep2023} to create creative interpretations of sound effects. In their approach, a desired sound-effect is described in text and embedded using CLAP, followed by iterative updates from a gradient-free optimizer~\cite{evosax2022github} to the synthesizer's parameters to minimize embedding differences. Based on manual hearing tests, this approach did not produce the ``correct'' sounds more frequently than previous works~\cite{kreuk2022audiogen}. However, it did yield better scores for ``artistic-interpretation'' (what we referred to as ``imitation''). It should be noted that audio embedding models are often trained using simpler loss functions---particularly those which utilize Fourier transforms---therefore improving such simpler loss functions is likely to improve the embedding models.


\subsection{What is Lacking In the Field}
\label{sec:lacking}

Having looked at the current literature, we identify four major areas of weakness in sound-matching. We target the first two issues in this work, while the latter two are left for future work.
\begin{enumerate}
    \item \LossSelect: It remains unclear whether there exists a universally best-performing loss function, 
    or whether performance depends on factors such as the sound domain, synthesizer architecture, 
    and desired output characteristics. Existing studies typically evaluate losses only with simple synthesis 
    setups~\cite{vahidi2023mesostructures}, leaving open questions about their generality. 
    \item \SynthSelect: There is a lack of diversity in the synthesis methods used in sound-matching. The effects of using different isolated DSP functions in iterative sound-matching and how these interact with loss functions remains largely untested.
    \item \PeriodicLoss: Loss function landscapes are not easy to navigate, as there can be many local minima, or large flat areas~\cite{turian2020sorry,vahidi2023mesostructures}.  In differentiable settings, this can cause gradient descent updates to not reach the global minima. 
    \item \OutDomain: Sound-matching with out-of-domain sounds is an under-explored area, yet a necessary one for practical applications for sound designers.
\end{enumerate}

\subsection{What Approach Should We Take?}
We argue that isolated benchmarking is a necessary precursor to building robust, generalizable sound-matching systems. To address the problems of \LossSelect{} and \SynthSelect{}, we adopt a methodology with three key characteristics that have not previously appeared together.

\begin{enumerate}
    \item Use simple differentiable synthesizers built from common DSP functions (oscillators, filters, envelopes) to address the \SynthSelect{} problem. Restriction to small-scale, low-parameter models avoids the confounding effects of neural proxies or embedding models, enabling direct analysis of how synthesis method interacts with loss functions. 
    \item Evaluate experiments under multiple loss functions to address the \LossSelect{} problem. Prior work has typically tested losses in isolation, making direct comparisons of loss functions difficult~\cite{vahidi2023mesostructures,han2023perceptual,uzrad2024diffmoog}. Particularly rare are comparisons which are done in the context of different methods of synthesis.  
    \item Optimize synthesizer parameters directly with gradient descent, without neural approximations. This complements prior proxy-based methods by providing the missing “low-level experiments” that clarify fundamental interactions between DSP functions and similarity measures. 
\end{enumerate}

Together, these design choices supply the controlled experimental evidence missing from the literature, 
clarifying how losses and synthesis methods interact in iterative sound-matching,
and offering insights likely to generalize to more complex domains.




\section{Experimental Setup and Results}
\label{sec:experiment_setup}
At a glance, our methodology is to conduct controlled, low-level differentiable experiments without the use of proxy networks. We pair four differentiable losses with four differentiable DSP-based synthesizers: BP-Noise (band-pass noise with LP/HP filters), Add-SineSaw (additive sine and saw oscillators), Noise-AM (noise modulated by an LFO), and SineSaw-AM (sine-modulated saw oscillator). We optimize parameters directly via gradients in order to isolate loss–synthesizer interactions. This controlled setup systematically evaluates the performance of iterative sound-matching pipelines, where “performance” is defined as the similarity of the synthesized output to a target. Similarity is assessed with P-Loss, MSS, and—most importantly—manual listening tests. This design allows us to directly test our central hypothesis: that the effectiveness of a loss function depends on the synthesis method, and that no universal “best” similarity measure exists.

Here we discuss the implementation of the four loss functions (Section~\ref{sec:loss_implementation})
across four differentiable synthesizers (Section~\ref{sec:programs}). For
each loss–synthesizer pair, we run a large number of trials and evaluate
their outcomes automatically and manually
(Section~\ref{sec:evaluation_manual_auto}). ``Experiment'' in this context
refers to one complete iterative sound-matching run, from random parameter
initialization through iterative gradient-based optimization until termination. Scores
from these experiments are then aggregated and compared to
determine best-performing losses and synth–loss pairings.

\subsection{Loss Function Implementation Details}
\label{sec:loss_implementation}
\subsubsection{STFT Losses}
Due to their ubiquity and lower cost of gradient calculation, we use STFT as the basis of two loss functions. 
L1 or L2 distances are the most common methods of comparing spectrograms~\cite{turian2020sorry,richard2025model}. 
However, these measures can be overly sensitive to global gain or minor misalignments, potentially overstating perceptual differences. 
To address this, we also test Scale-Invariant Mean Squared Error (SIMSE) as an alternative to L1 for comparing STFT spectrograms.

We define the \LoneSpec~and \SIMSESpec~functions as the application of L1 and SIMSE to the STFT spectrograms. The STFT spectrogram function uses 512 FFT bins, window size of 600 samples, and hop length (how many samples the window shifts) of 100 samples. The L1 and SIMSE implementations are differentiable. 

\subsubsection{JTFS Loss}
The \JTFS~loss function is the application of L1 difference to the JTFS representations of two sounds~\cite{vahidi2023mesostructures}. The code used for the JTFS transformation is the differentiable implementation of a 1-dimensional JTFS function provided by Andreux \textit{et al.}~\cite{kymatio}. 

\subsubsection{Soft-DTW Loss}
We use the soft-DTW function, which is differentiable and---depending on its parameters---not shift-invariant~\cite{cuturi2017soft,janati2020spatio,tavenard.blog.softdtw}. The loss function \DTWEnv~is the application of the soft-DTW function to the envelope of the two sounds being compared~\cite{lyons1997understanding}. This loss uses the similarity of amplitude modulation patterns, which are perceptually important in many sound-design scenarios. The envelope is calculated by creating the STFT spectrogram of a sound (the same process used in Section~\ref{sec:fourier_specs}) and summing the values at each timestep. 

\subsection{The Synthesizers}
\label{sec:programs}
The synthesizer programs are meant to be simple examples that test the building blocks of digital sound synthesis. Subtractive, additive, and FM/AM synthesis are three of the most common techniques in sound design~\cite{smith1991viewpoints}. In subtractive synthesis, frequencies are removed from a sound via digital filters. In additive synthesis, complex sounds are created via the linear combination of simpler sounds~\cite{lyons1997understanding,smith2007introduction}. As discussed previously, FM/AM synthesis refers to the general technique of modulating the frequency or amplitude of a waveform (the carrier) by another waveform (the modulator).

For each program, we provide the Faust code~\cite{orlarey2009faust}, which can be run in the online IDE~\footnote{\url{https://faustide.grame.fr/}}. Faust is a functional language for audio synthesis that can succinctly define signal processing chains. Following Braun's methodology~\cite{braun2024dac}, programs are first defined in Faust, then converted (or \textit{transpiled}) to differentiable JAX functions using the DawDreamer library\footnote{\url{https://github.com/DBraun/DawDreamer}}. 
\begin{lstlisting}[caption={\BPNoise}, label={lst:program0}, language=Faust,
                  float, floatplacement=!H, xleftmargin=1em, xrightmargin=0.5em, firstnumber=0, aboveskip=0em, belowskip=-1em]
import("stdfaust.lib");
lp_cut = hslider("lp_cut",900,50,1000,1);
hp_cut = hslider("hp_cut",100,1,120,1);
process = no.noise:fi.lowpass(3,lp_cut):fi.highpass(10,hp_cut);
\end{lstlisting}

\begin{lstlisting}[caption={\AddSineSaw}, label={lst:program1},language=Faust,float,floatplacement=!H,xleftmargin=1em,xrightmargin=0.5em,firstnumber=0,aboveskip=0em, belowskip=-1em]
import("stdfaust.lib");
saw_freq = hslider("saw_freq",800,20,1000,1);
sine_freq = hslider("sine_freq",300,20,1000,1);
sineOsc(f) = +(f/ma.SR) ~ ma.frac:*(2*ma.PI) : sin;
sawOsc(f) = +(f/ma.SR) ~ ma.frac;
process = sineOsc(sine_freq)+sawOsc(saw_freq);
\end{lstlisting}

\begin{lstlisting}[caption={\AmpMod}, label={lst:program2},language=Faust,float,floatplacement=!H,xleftmargin=1em,xrightmargin=0.5em,firstnumber=0,aboveskip=0em, belowskip=-1em]
import("stdfaust.lib");
amp = hslider("amp",0.5,0,5,0.01);
modulator = hslider("modulator",0.5,0,4,0.01);
sineOsc(f) = +(f/ma.SR) ~ ma.frac:*(2*ma.PI) : sin;
process = no.noise*sineOsc(modulator)*amp;
\end{lstlisting}

\begin{lstlisting}[caption={\FMMod}, label={lst:program3},language=Faust,float,floatplacement=!H,xleftmargin=1em,xrightmargin=0.5em,firstnumber=0,aboveskip=0em, belowskip=-1em]
import("stdfaust.lib");
carrier = hslider("carrier",100,20,1000,1);
amp = hslider("amp",6,1,20,1);
sineOsc(f) = +(f/ma.SR) ~ ma.frac:*(2*ma.PI) : sin;
sawOsc(f) = +(f/ma.SR) ~ ma.frac;
process = sineOsc(amp)*sawOsc(carrier);
\end{lstlisting}

\subsubsection{\BPNoise}
\label{sec:program0}
\BPNoise{} is a bare bones example of subtractive synthesis using digital filters~\cite{smith2007introduction}. Despite the ubiquity of subtractive synthesis in practical sound design, it has rarely been tested in differentiable sound-matching. In this program, a noise signal is fed through a band-pass (BP) filter. This removes the frequencies outside the low-pass (LP) and high-pass (HP) cutoffs. The noise generator produces all frequencies with random variation over time. The LP filter removes frequencies over its threshold frequency, and the HP removes frequencies lower than its threshold~\cite{smith2007introduction}. The search parameters are the cutoff thresholds for the LP and HP filters. Listing~\ref{lst:program0} shows the Faust code for \BPNoise.

\subsubsection{\AddSineSaw}
\label{sec:program1}
\AddSineSaw{} is an additive program that combines a saw and a sine wave function. Like subtractive synthesis, additive synthesis is a common approach to sound design that has not been extensively tested as a benchmark in sound-matching. The search parameters here are the frequency of the sine and saw oscillators. Listing~\ref{lst:program1} is the Faust code for \AddSineSaw.

\subsubsection{\AmpMod}
\label{sec:program2}
\AmpMod{} involves amplitude modulation of a noise generator by an LFO with \textit{modulator} as its frequency and a global \textit{amp} value that does not change over time and applies to the entire signal. The search parameters for this program are the LFO frequency and the amp value. The amp value acts as a global volume control, and since the sounds are normalized before analysis, it is likely inconsequential to the performance. This program is meant to be a stepping stone to \FMMod, which utilizes proper AM synthesis. Listing~\ref{lst:program2} is the Faust code for \AmpMod. 

\subsubsection{\FMMod}
\label{sec:program3}
Relative to additive and subtractive synthesis, AM/FM synthesis are less common methods of sound design. Due to their frequent use in previous works, we also tested an AM synthesizer. The synthesizer program is the multiplication of a low frequency sine oscillator with frequency parameter \textit{amp}, and a saw oscillator with frequency parameter \textit{carrier}. Listing~\ref{lst:program3} is the Faust code for \FMMod.

\subsection{Evaluation Methods}
\label{sec:evaluation_manual_auto}
The two automatic evaluation methods used here are P-Loss and MSS, as described in Section~\ref{sec:loss_funcs}. For P-Loss, the parameters are normalized between 0-1 based on the valid ranges defined in the Faust program, and the L1 distance between the normalized parameters is calculated. MSS is computed using a hop size of $100$ samples, and FFT window lengths of $(512, 1024, 2048, 4096)$. 

Automatic sound evaluation methods are often unreliable, as their alignment with the subjective judgments of human listeners is uncertain. Therefore, listening tests are conducted by randomly sampling 40 experiment results for each program and loss function pair. Two of the authors then assign blinded similarity scores to the outputs and targets using a 5-point Likert scale (1 = no similarity, 5 = near identical)~\cite{jebb2021review}. Using Spearman's rank correlation~\cite{spearman1987proof,rebekic2015pearson}, we found a very strong rate of agreement between the human listeners (see Section~\ref{sec:consistency_in_rankings}), a good indicator that the scores assigned by manual listeners can be treated as the ground truth. We then combine the ranks assigned by both authors, giving us 80 ranks per program and loss function pair.



\subsection{Training Loop and Gradient Calculations}
Given a differentiable loss and synthesizer, the iterative sound-matching procedure is as follows:
 \begin{enumerate}
    \item Initialize $\theta^*$ and $\hat{\theta}$, i.e., random generation of target and initial parameters uniformly over a predefined range
    \item Generating the output of the synthesizer with $\hat{\theta}$ (the length of the output is set to 1 second with sample-rate of 44100 Hz) 
    \item Calculating the loss between the target and output
    \item Applying gradient updates to the synthesizer parameters
    \item Repeating the second step with the updated parameters $\hat{\theta}$, until maximum number of iterations has been reached
 \end{enumerate}

 For updating the synthesizer parameters (or weights), we use RMSProp, which operates similar to stochastic gradient descent (SGD)~\cite{goodfellow2016deep}, with the caveat that the gradients of each weight are scaled by the root-mean-square of past gradients of that weight. Based on some initial test runs, we used an arbitrary fixed learning rate of 0.045 for all experiments (learning rates used in Vahidi \textit{et al.}~\cite{vahidi2023mesostructures} is unknown). The maximum number of iterations is set to 200, where, based on our observations, the parameters have either irrecoverably diverged outside the acceptable ranges, or are stuck at a local minimum. Additionally, the gradients are large, and calculated backwards sequentially through the signal processing chain. This backward calculation resembles the \textit{exploding gradients} problem in recurrent neural networks~\cite{gers2000learning}. Gradient clipping~\cite{goodfellow2016deep} is preemptively used to ensure that the  $\ell_2$ norm of all gradients does not exceed the threshold of 1. 
\subsection{Ranking Loss Functions}

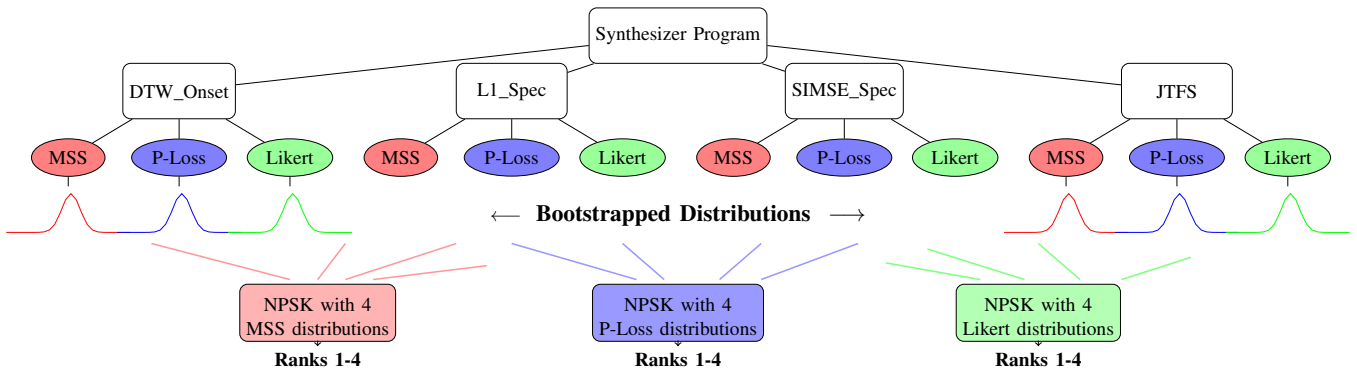
\begin{figure*}
\resizebox{\linewidth}{!}{ 
\begin{tikzpicture}[
    function/.style={rectangle, draw, rounded corners, minimum width=2cm, minimum height=1cm, text centered, node distance=1cm},
    distribution/.style={ellipse, draw, minimum width=1cm, minimum height=0cm, text centered, node distance=0.1cm},
    gaussian/.style={draw, smooth, samples=100, domain=-2.5:2.5},
    NPSK/.style={rectangle, draw, rounded corners, minimum width=2.5cm, minimum height=1cm, text centered, node distance=1.5cm, fill=blue!20},
    level 1/.style={sibling distance=6cm, level distance=1cm},
    level 2/.style={sibling distance=2cm, level distance=1.2cm},
    mss/.style={fill=red!50, draw=black},
    ploss/.style={fill=blue!50, draw=black},
    likert/.style={fill=green!40, draw=black}
]

\node[function] {Synthesizer Program}
    child { node[function] {DTW\_Onset}
        child { node[distribution, mss] {MSS}
            child[grow=down, level distance=1cm] {
                node[draw=none] {
                    \begin{tikzpicture}[scale=0.7]
                        \draw[red] plot (\x/3, {exp(-\x*\x)});  
                    \end{tikzpicture}
                }
            }
        }
        child { node[distribution, ploss] {P-Loss}
            child[grow=down, level distance=1cm] {
                node[draw=none] {
                    \begin{tikzpicture}[scale=0.7]
                        \draw[blue] plot (\x/3, {exp(-\x*\x)});  
                    \end{tikzpicture}
                }
            }
        }
        child { node[distribution, likert] {Likert}
            child[grow=down, level distance=1cm] {
                node[draw=none] {
                    \begin{tikzpicture}[scale=0.7]
                        \draw[green] plot (\x/3, {exp(-\x*\x)});  
                    \end{tikzpicture}
                }
            }
        }
    }
    child { node[function] {L1\_Spec}
        child { node[distribution, mss] {MSS} }
        child { node[distribution, ploss] {P-Loss} }
        child { node[distribution, likert] {Likert} }
    }
    child { node[function] {SIMSE\_Spec}
        child { node[distribution, mss] {MSS} }
        child { node[distribution, ploss] {P-Loss} }
        child { node[distribution, likert] {Likert} }
    }
    child { node[function] {JTFS}
        child { node[distribution, mss] {MSS}
            child[grow=down, level distance=1cm] {
                node[draw=none] {
                    \begin{tikzpicture}[scale=0.7]
                        \draw[red] plot (\x/3, {exp(-\x*\x)});  
                    \end{tikzpicture}
                }
            }
        }
        child { node[distribution, ploss] {P-Loss}
            child[grow=down, level distance=1cm] {
                node[draw=none] {
                    \begin{tikzpicture}[scale=0.7]
                        \draw[blue] plot (\x/3, {exp(-\x*\x)});  
                    \end{tikzpicture}
                }
            }
        }
        child { node[distribution, likert] {Likert}
            child[grow=down, level distance=1cm] {
                node[draw=none] {
                    \begin{tikzpicture}[scale=0.7]
                        \draw[green] plot (\x/3, {exp(-\x*\x)});  
                    \end{tikzpicture}
                }
            }
        }
    };

\node at (0,-3.25) {\textbf{$\longleftarrow$\ \ \large Bootstrapped Distributions\ \ $\longrightarrow$}};

\def\xoffsetL{-0.5}
\def\xoffsetM{0}
\def\xoffsetR{0.5}
\def\yoffset{1}

\node[NPSK, text height=5ex, align=center, fill=red!30, draw=black] at ({-6+\xoffsetL},{-6+\yoffset}) {NPSK with 4 \\ MSS distributions};
\draw[->] ({-6+\xoffsetL},{-6.5+\yoffset}) -- ({-6+\xoffsetL},{-6.6+\yoffset}) node[below, below] {\textbf{Ranks 1-4}};
\draw[red!40, thick] ({-5+\xoffsetL}, {-5.4+\yoffset}) -- ({-2.95+\xoffsetL}, {-5.15+\yoffset});  
\draw[red!40, thick] ({-5.5+\xoffsetL}, {-5.4+\yoffset}) -- ({-3.5+\xoffsetL}, {-4.75+\yoffset});  
\draw[red!40, thick] ({-6+\xoffsetL}, {-5.4+\yoffset}) -- ({-5.5+\xoffsetL}, {-4.75+\yoffset});  
\draw[red!40, thick] ({-6.5+\xoffsetL}, {-5.4+\yoffset}) -- ({-9+\xoffsetL}, {-4.75+\yoffset});  

\node[NPSK, text height=5ex, align=center, fill=blue!40, draw=black] at ({0+\xoffsetM},{-6+\yoffset}) {NPSK with 4 \\ P-Loss distributions};
\draw[->] ({0+\xoffsetM},{-6.5+\yoffset}) -- ({0+\xoffsetM},{-6.6+\yoffset}) node[below, below] {\textbf{Ranks 1-4}};
\draw[blue!40, thick] ({1.5+\xoffsetM}, {-5.4+\yoffset}) -- ({3.25+\xoffsetM}, {-4.75+\yoffset});  
\draw[blue!40, thick] ({0.75+\xoffsetM}, {-5.4+\yoffset}) -- ({1.5+\xoffsetM}, {-4.75+\yoffset});  
\draw[blue!40, thick] ({-0.25+\xoffsetM}, {-5.4+\yoffset}) -- ({-1+\xoffsetM}, {-4.75+\yoffset});  
\draw[blue!40, thick] ({-0.75+\xoffsetM}, {-5.4+\yoffset}) -- ({-3+\xoffsetM}, {-4.75+\yoffset});  

\node[NPSK, text height=5ex, align=center, fill=green!30, draw=black] at ({6+\xoffsetR},{-6+\yoffset}) {NPSK with 4 \\ Likert distributions};
\draw[->] ({6+\xoffsetR},{-6.5+\yoffset}) -- ({6+\xoffsetR},{-6.6+\yoffset}) node[below, below] {\textbf{Ranks 1-4}};
\draw[green!50, thick] ({7.5+\xoffsetR}, {-5.4+\yoffset}) -- ({8.75+\xoffsetR}, {-5+\yoffset});  
\draw[green!50, thick] ({6.75+\xoffsetR}, {-5.4+\yoffset}) -- ({6+\xoffsetR}, {-4.75+\yoffset});    
\draw[green!50, thick] ({5.75+\xoffsetR}, {-5.4+\yoffset}) -- ({4+\xoffsetR}, {-4.85+\yoffset});    
\draw[green!50, thick] ({4.95+\xoffsetR}, {-5.4+\yoffset}) -- ({3.25+\xoffsetR}, {-5.1+\yoffset}); 

\end{tikzpicture}
}
\caption{For each synthesizer program, we assign ranks to the four loss functions  (\DTWEnv, \LoneSpec, \SIMSESpec, \JTFS) in three different ways, using MSS, P-Loss, or manually assigned Likert scores.}
\label{fig:posthoc_evaluation}
\end{figure*}

 
 Figure~\ref{fig:posthoc_evaluation} is a visualized summary of how the loss functions are ranked. We have four loss functions and four synthesizer programs. The maximum number of iterations set to 200, and 300 experiments for each loss/synthesizer combination. Each synthesizer program is paired with four loss functions and 300 experiments are conducted and evaluated automatically with P-Loss and MSS, and manually with Likert scores. With this approach, for each program and loss function pair, there are three distributions that can be used to rank the loss functions. Two distributions with 300 automatically assigned similarities (P-Loss and MSS) and one distribution with 80 Likert scores (combining the 40 ranks assigned by each author).

For consistency and statistical robustness, the distributions are upsampled to 1000 values using bootstrapping~\cite{tibshirani1993introduction}. Bootstrapping gives an estimation of the distribution for the mean performance of each experiment. $k$ (set to 1000) samples of $n$ values (set to 100\% of the values) are taken with replacement from the empirical distribution (list of 300 performance values for MSS and P-Loss or 80 for manual rankings). The mean performance is calculated for each of the $k$ samples. The $k$ estimates of the mean performance give us a bootstrapped distribution for each loss function.

 A post-hoc test is conducted in two stages in order to determine whether there is a difference in performance between loss functions for each program, and if so, which are the best performers. The first stage determines whether there is a difference in group means, with the null hypothesis being that all groups have similar mean ranks. The second stage ranks the loss functions from best to worst using the non-parametric Scott-Knott test (\gls{NPSK})~\cite{tantithamthavorn2017mvt,tantithamthavorn2018optimization}.
The Kruskal-Wallis test is used for the first stage~\cite{kruskal1952use}; this test pools and ranks all evaluation measures for a program, then tests whether the differences in mean rank for all loss function groups is zero. The Kruskal-Wallis test calculates an H statistic that is compared to a chi-square distribution with k-1 degrees of freedom (k is the number of groups, or 4, and degrees of freedom is 3). Using the significance level of 0.05, if the H statistic is greater than the critical value of the chi-square distribution then the null hypothesis is rejected, meaning that at least one group has a mean rank significantly above others~\cite{kruskal1952use}. 

\begin{table}[ht]
\centering
\caption{Do loss functions have different median performance? Kruskal-Wallis results by program and bootstrapped evaluation results.}
\begin{tabular}{|c|c|c|c|c|}
\hline
\textbf{Program} & \textbf{Eval. Method} & \textbf{H-Stat.} & \textbf{P-value} & \textbf{Reject} \\
\hline
\BPNoise & MSS      & 81.19  & $1.70 \times 10^{-17}$ & Yes \\
\BPNoise & P-Loss  & 163.60 & $3.07 \times 10^{-35}$ & Yes \\
\BPNoise & Manual & 9.45  & $2.39 \times 10^{-2}$ & Yes \\
\AddSineSaw & MSS      & 308.97 & $1.14 \times 10^{-66}$ & Yes \\
\AddSineSaw & P-Loss  & 348.42 & $3.28 \times 10^{-75}$ & Yes \\
\AddSineSaw & Manual & 9.45  & $2.39 \times 10^{-2}$ & Yes \\
\AmpMod & MSS      & 1.35   & $0.7171$               & No \\
\AmpMod & P-Loss  & 366.76 & $3.50 \times 10^{-79}$ & Yes \\
\AmpMod & Manual & 32.71 & $3.70 \times 10^{-7}$ & Yes \\
\FMMod & MSS      & 564.65 & $4.65 \times 10^{-122}$ & Yes \\
\FMMod & P-Loss  & 229.19 & $2.07 \times 10^{-49}$ & Yes \\
\FMMod & Manual &  207.58 & $9.69 \times 10^{-45}$ & Yes \\
\hline
\end{tabular}
\label{tab:kruskal_auto}
\end{table}


\subsection{Best Performers For Each Program}
Table~\ref{tab:kruskal_auto} shows the Kruskal-Wallis results for every program and evaluation method, we see significant differences between loss function performance measures (i.e., the bootstrapped distributions of the evaluation for each loss function) in 11 of the 12 cases, with the exception of  MSS for the \AmpMod{} program.  

Based on the bootstrapped distribution of the scores, the NPSK algorithm ranks the loss functions from 1 (best) to a maximum of 4 (worst). In cases where the distributions are similar, multiple loss functions can be clustered into the same rank.

\begin{figure*}[p]
  \centering
\subfloat[\BPNoise~bootstrapped distributions and ranks given by NPSK.]{
  \begin{minipage}{\textwidth}
    \begin{minipage}[t]{0.03\textwidth}
      \footnotesize\raggedleft
      \vspace{0.65cm}
      SIMSE\\[0.6cm]
      L1\\[0.65cm]
      JTFS\\[0.65cm]
      DTW
    \end{minipage}%
    \hspace{0.01\textwidth}%
    \begin{minipage}[t]{0.96\textwidth}
      \centering
      \begin{minipage}[t]{0.31\textwidth}
        \centering
        \normalsize\bfseries MSS\\[0.3em]
        \includegraphics[width=\linewidth]{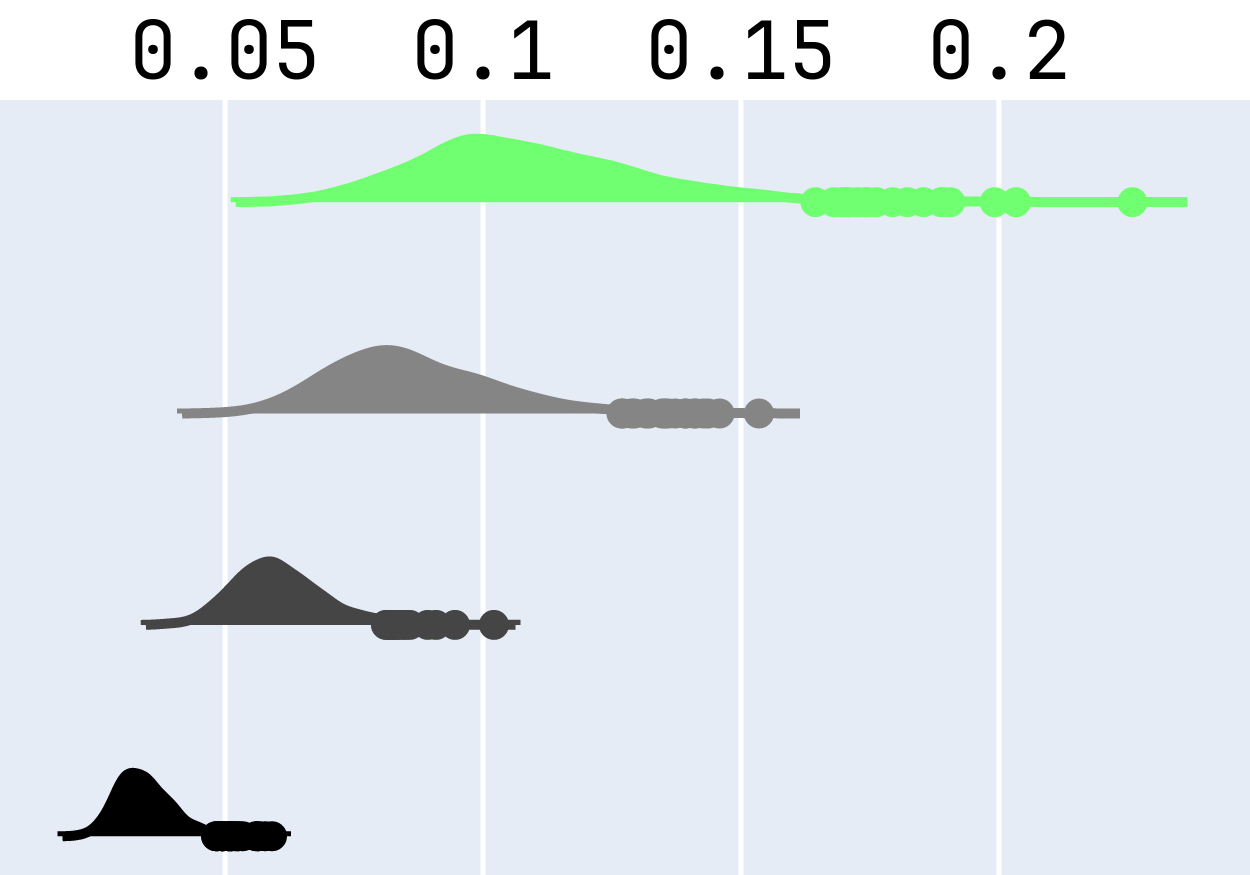}
      \end{minipage}
      \hspace{0.015\textwidth}%
      \begin{minipage}[t]{0.31\textwidth}
        \centering
        \normalsize\bfseries P-Loss\\[0.3em]
        \includegraphics[width=\linewidth]{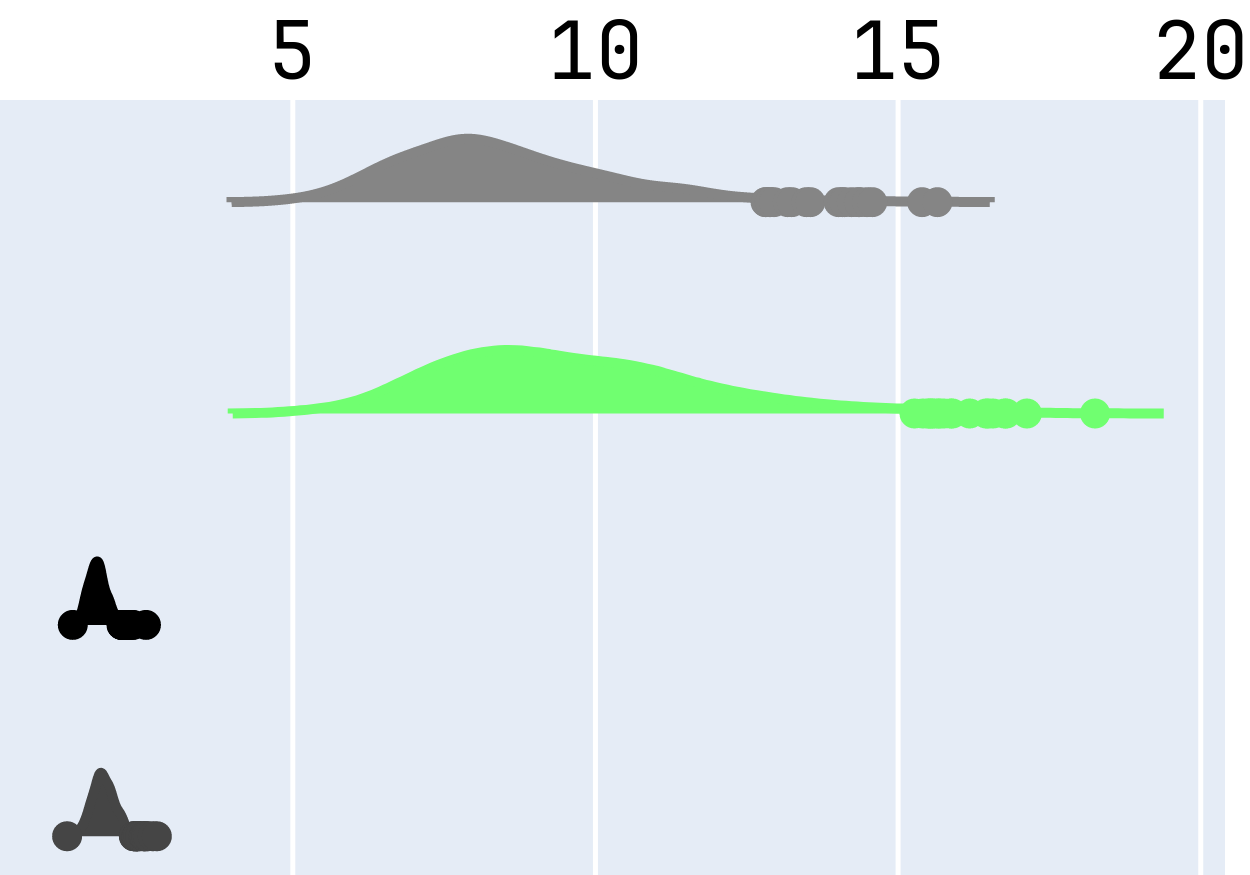}
      \end{minipage}
      \hspace{0.01\textwidth}%
      \begin{minipage}[t]{0.31\textwidth}
        \centering
        \normalsize\bfseries Likert\\[0.3em]
        \includegraphics[width=\linewidth]{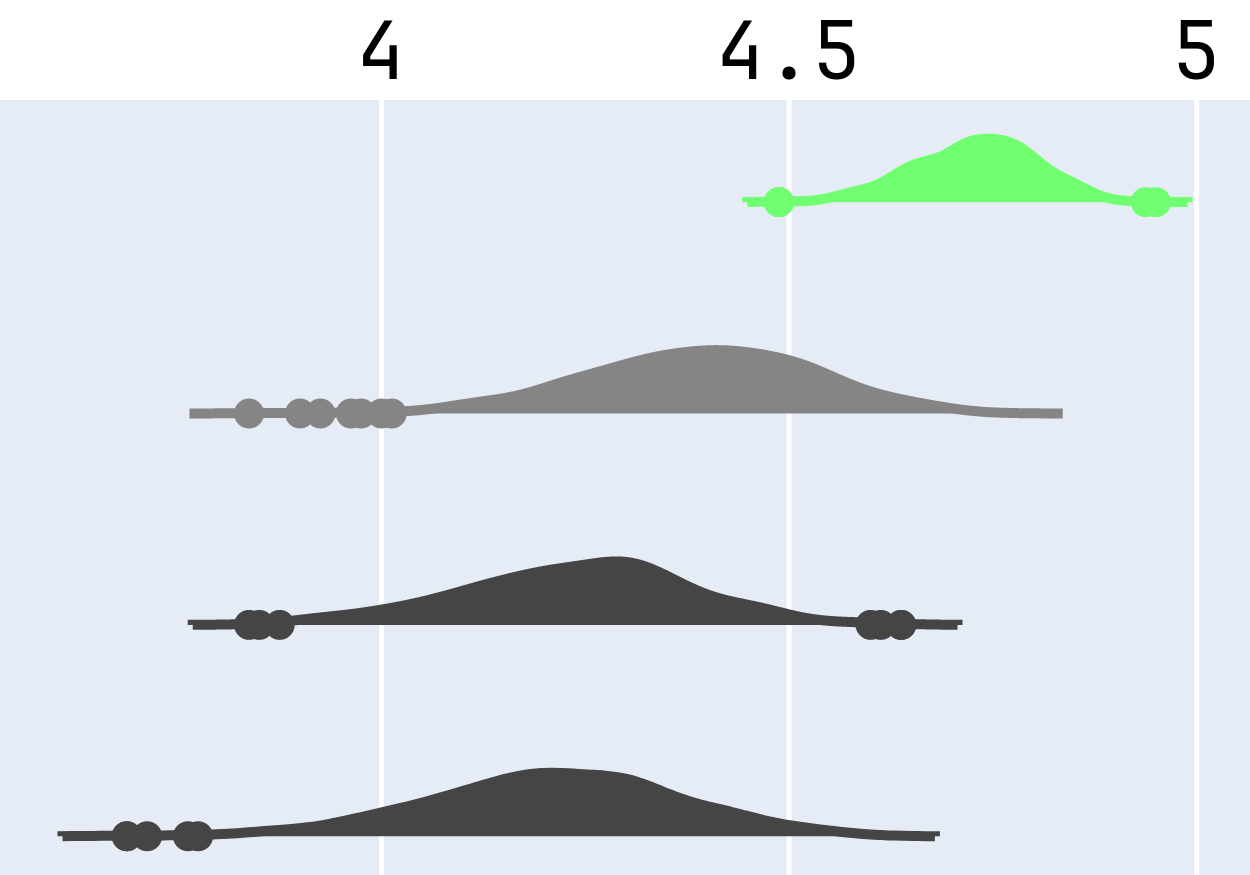}
      \end{minipage}
    \end{minipage}
  \end{minipage}
  \label{fig:npsk_p0}
}\\[0.5em]

\subfloat[\AddSineSaw~bootstrapped distributions and ranks given by NPSK.]{
  \begin{minipage}{\textwidth}
    \begin{minipage}{0.03\textwidth}
      \footnotesize\raggedleft
      \vspace{0.5cm}
      SIMSE\\[0.6cm]
      L1\\[0.65cm]
      JTFS\\[0.65cm]
      DTW
    \end{minipage}%
    \begin{minipage}{0.98\textwidth}\centering
      \includegraphics[width=0.31\textwidth]{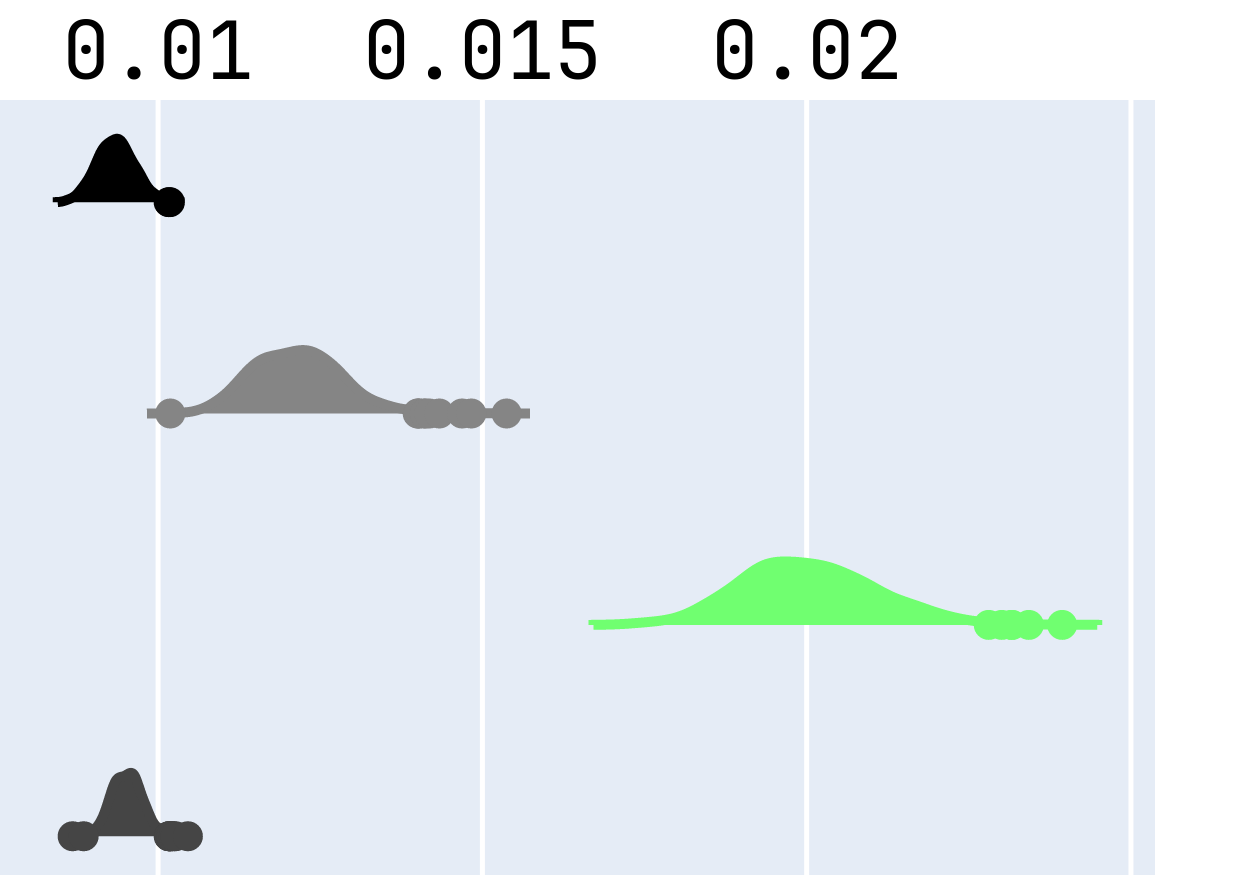}%
      \hspace{0.015\textwidth}%
      \includegraphics[width=0.31\textwidth]{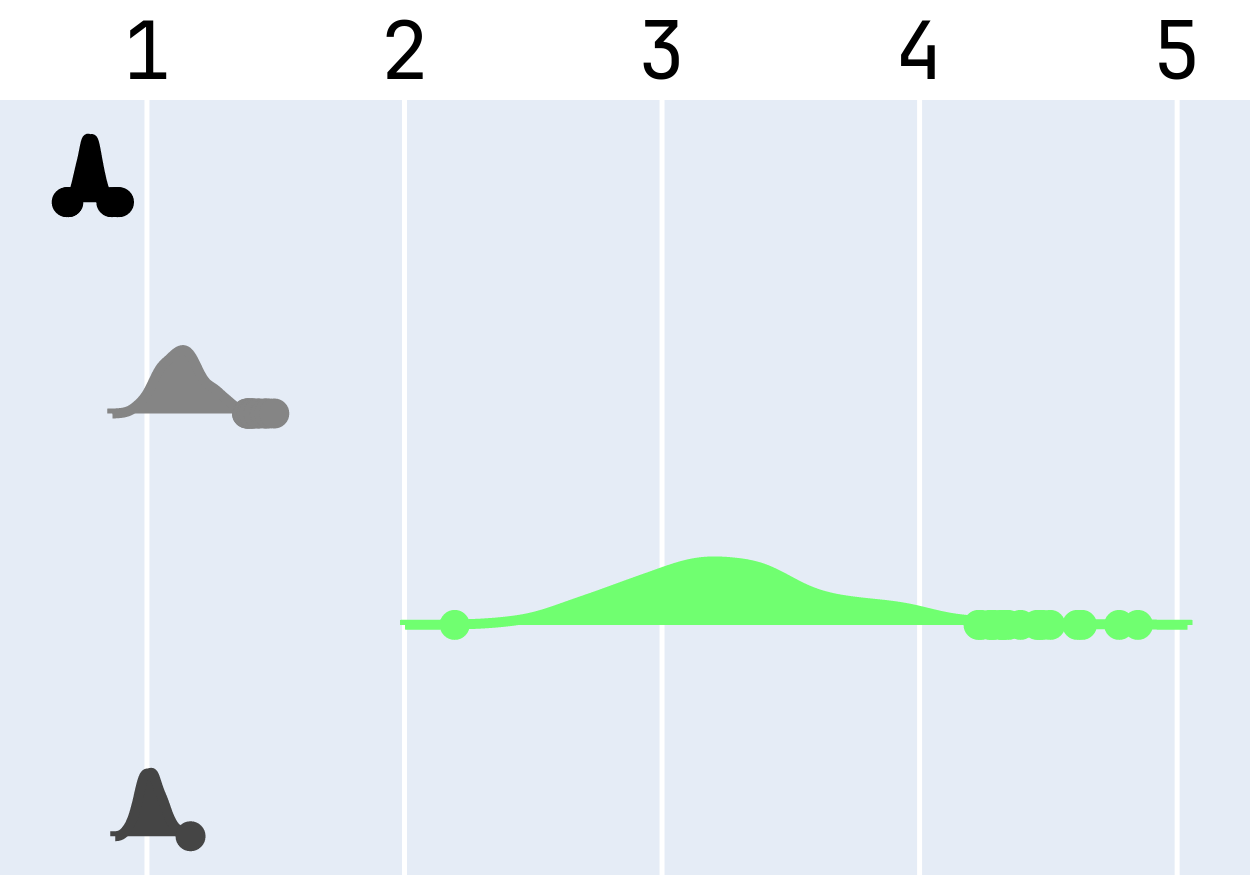}%
      \hspace{0.015\textwidth}%
      \includegraphics[width=0.31\textwidth]{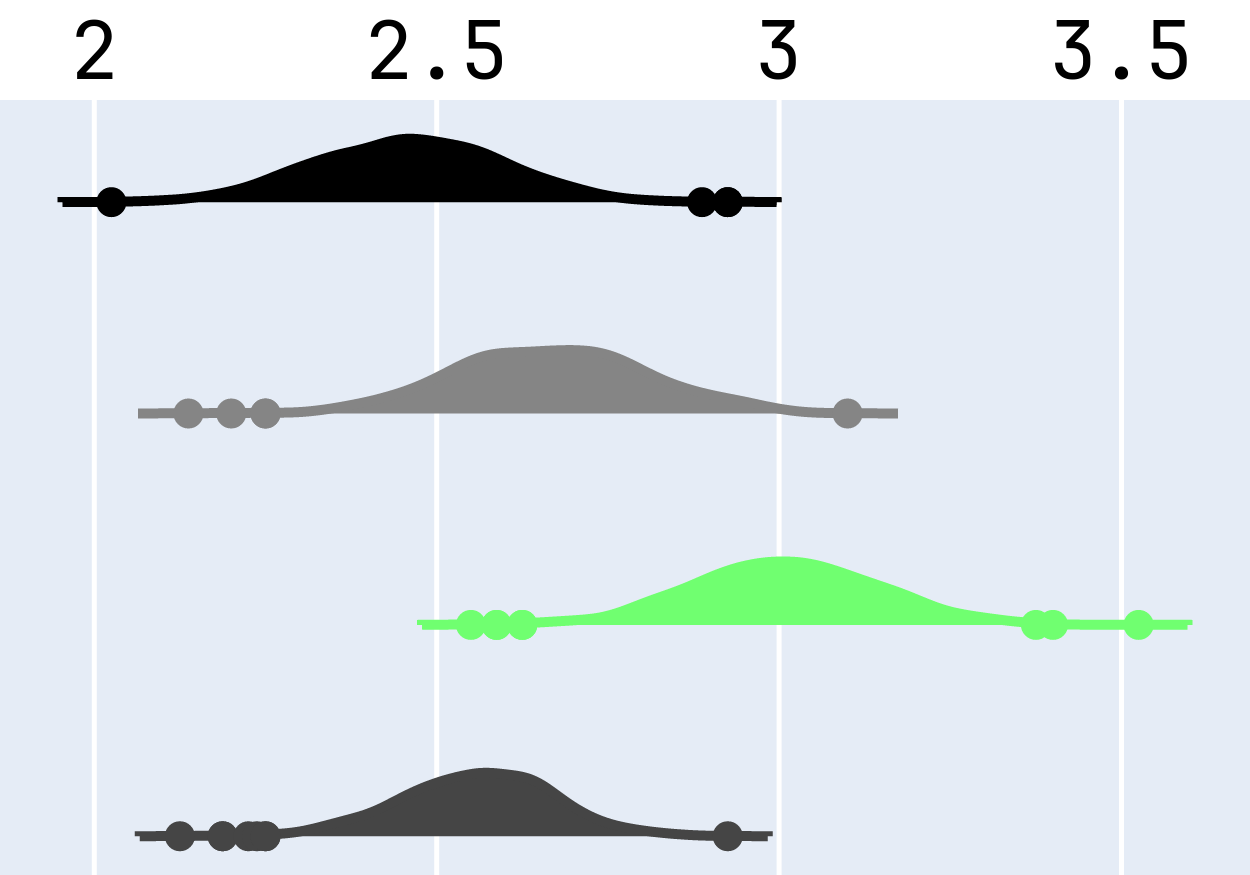}
    \end{minipage}
  \end{minipage}
  \label{fig:npsk_p1}
}\\[0.5em]

\subfloat[\AmpMod~bootstrapped distributions and ranks given by NPSK.]{
  \begin{minipage}{\textwidth}
    \begin{minipage}{0.03\textwidth}
      \footnotesize\raggedleft
      \vspace{0.5cm}
      SIMSE\\[0.6cm]
      L1\\[0.65cm]
      JTFS\\[0.65cm]
      DTW
    \end{minipage}%
    \begin{minipage}{0.98\textwidth}\centering
      \includegraphics[width=0.31\textwidth]{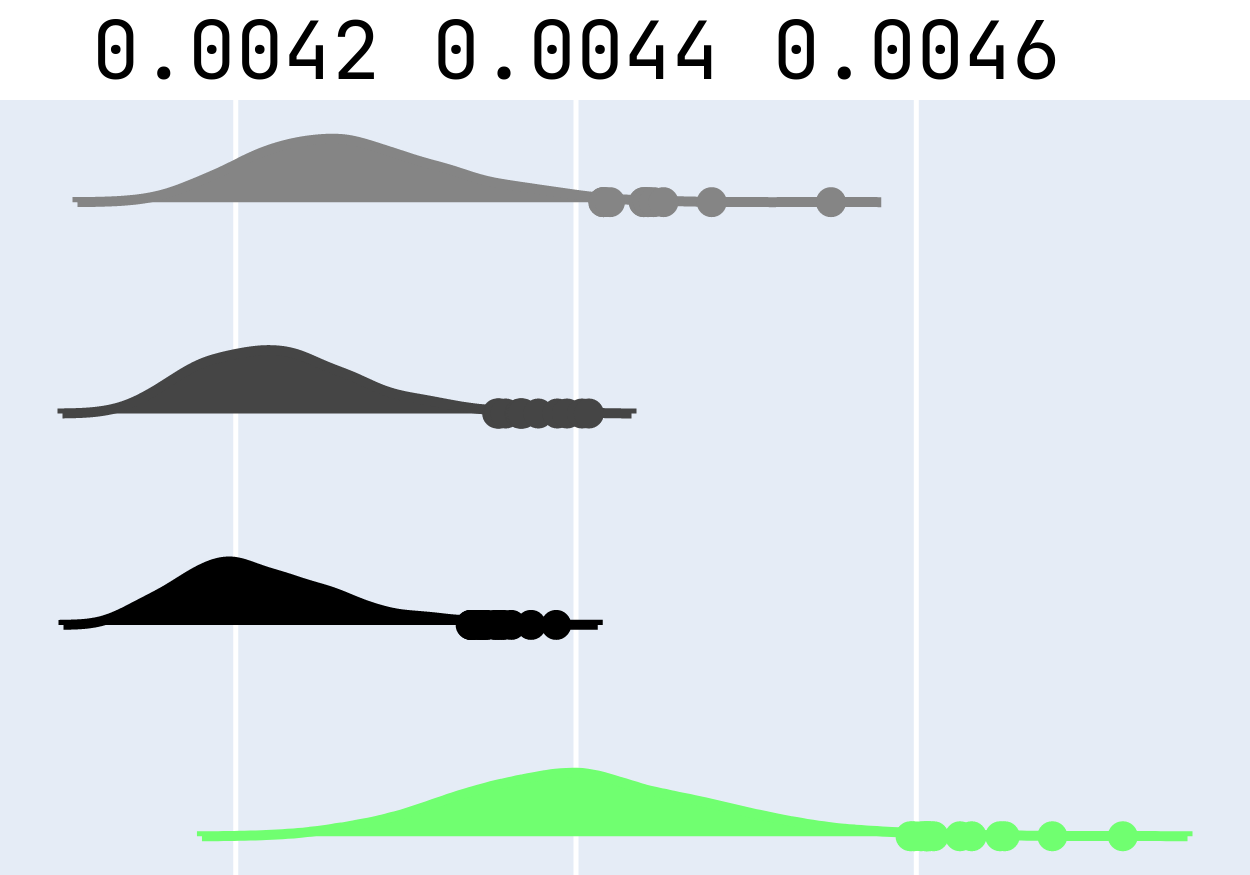}%
      \hspace{0.015\textwidth}%
      \includegraphics[width=0.31\textwidth]{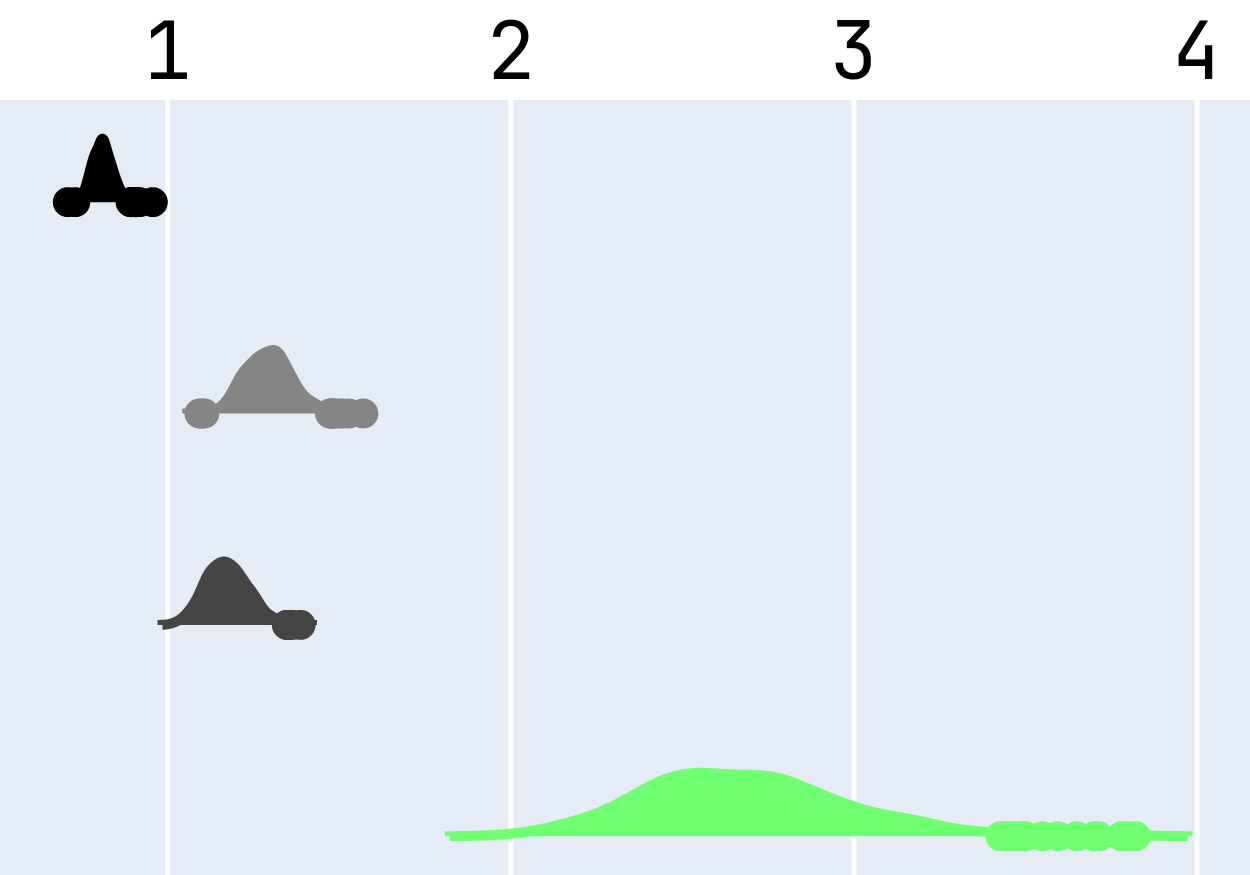}%
      \hspace{0.015\textwidth}%
      \includegraphics[width=0.31\textwidth]{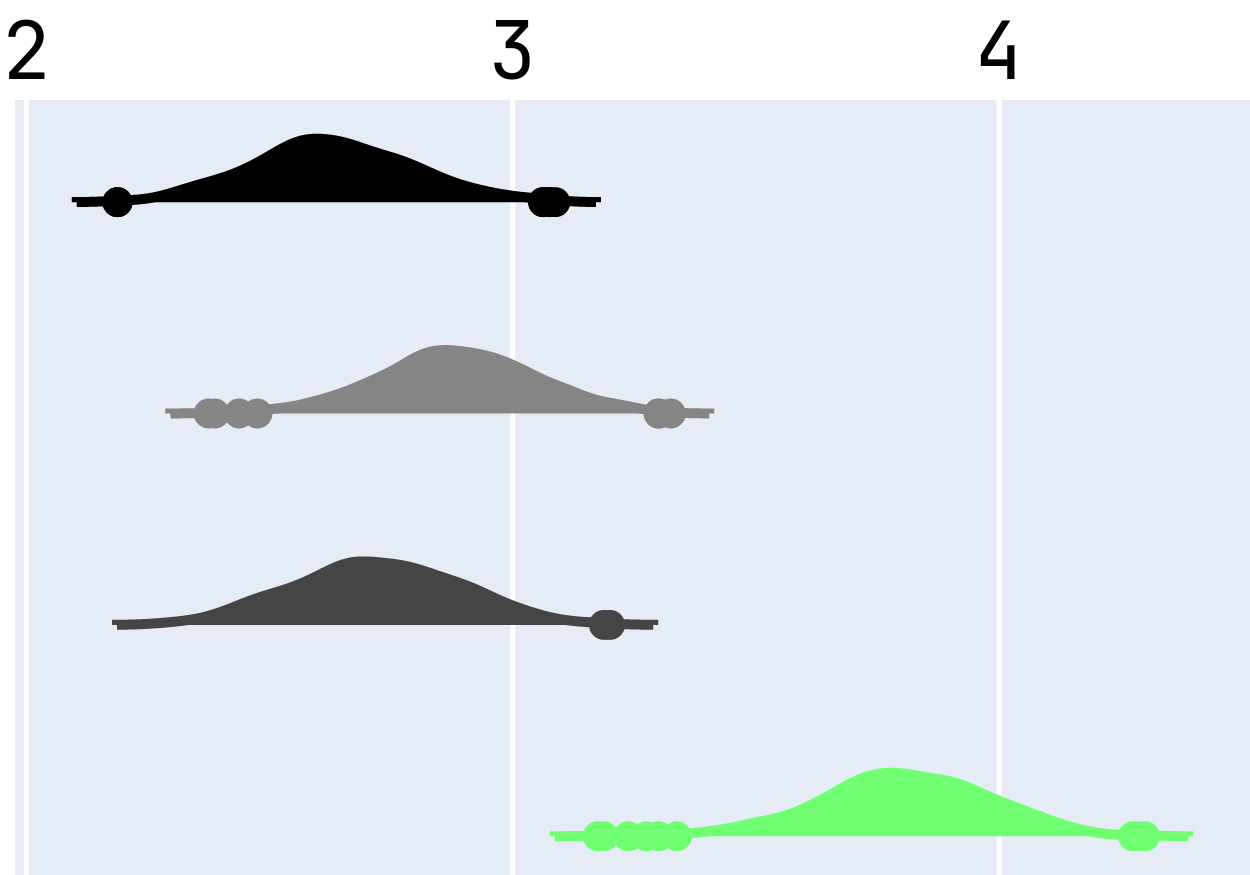}
    \end{minipage}
  \end{minipage}
  \label{fig:npsk_p2}
}\\[0.5em]

\subfloat[\FMMod~bootstrapped distributions and ranks given by NPSK.]{
  \begin{minipage}{\textwidth}
    \begin{minipage}{0.03\textwidth}
      \footnotesize\raggedleft
      \vspace{0.5cm}
      SIMSE\\[0.6cm]
      L1\\[0.65cm]
      JTFS\\[0.65cm]
      DTW
    \end{minipage}%
    \begin{minipage}{0.98\textwidth}\centering
      \includegraphics[width=0.31\textwidth]{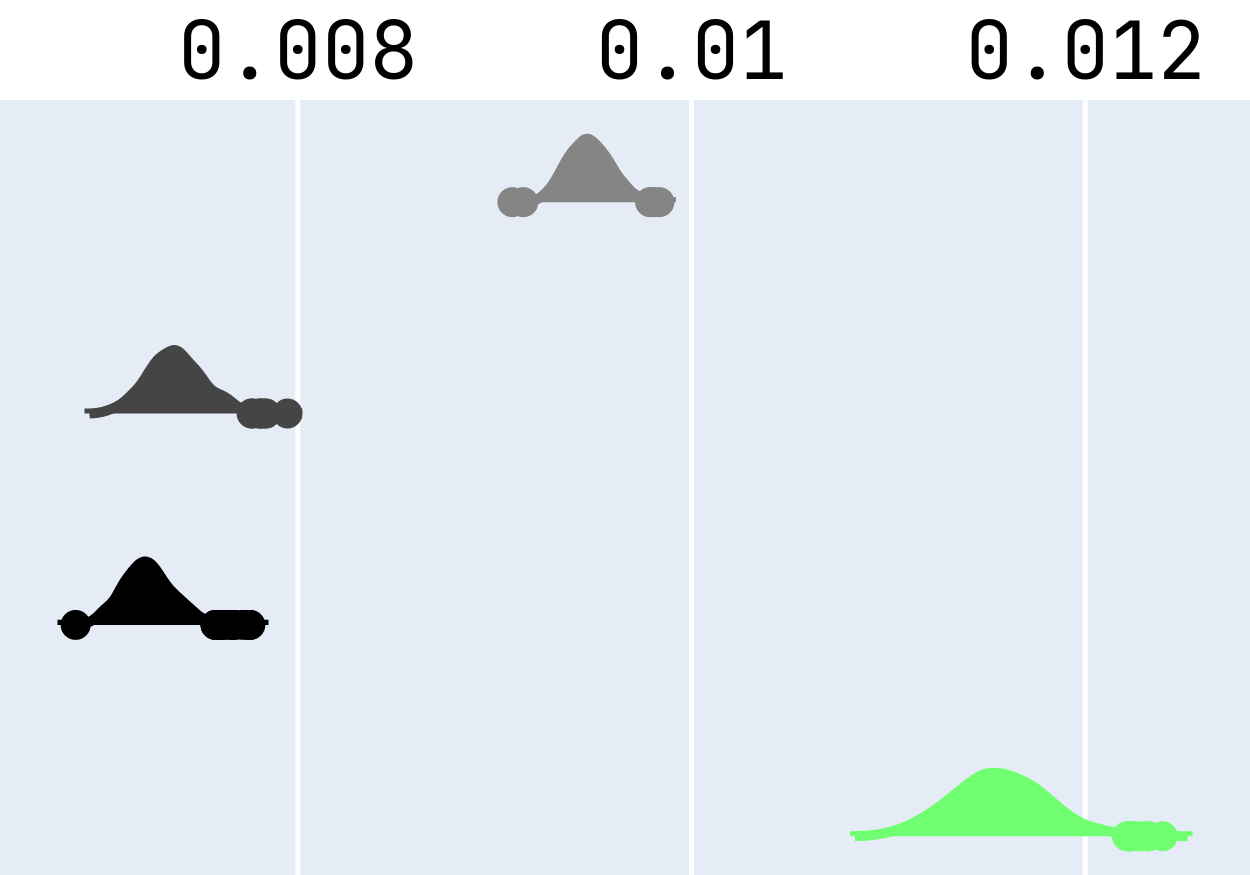}%
      \hspace{0.015\textwidth}%
      \includegraphics[width=0.31\textwidth]{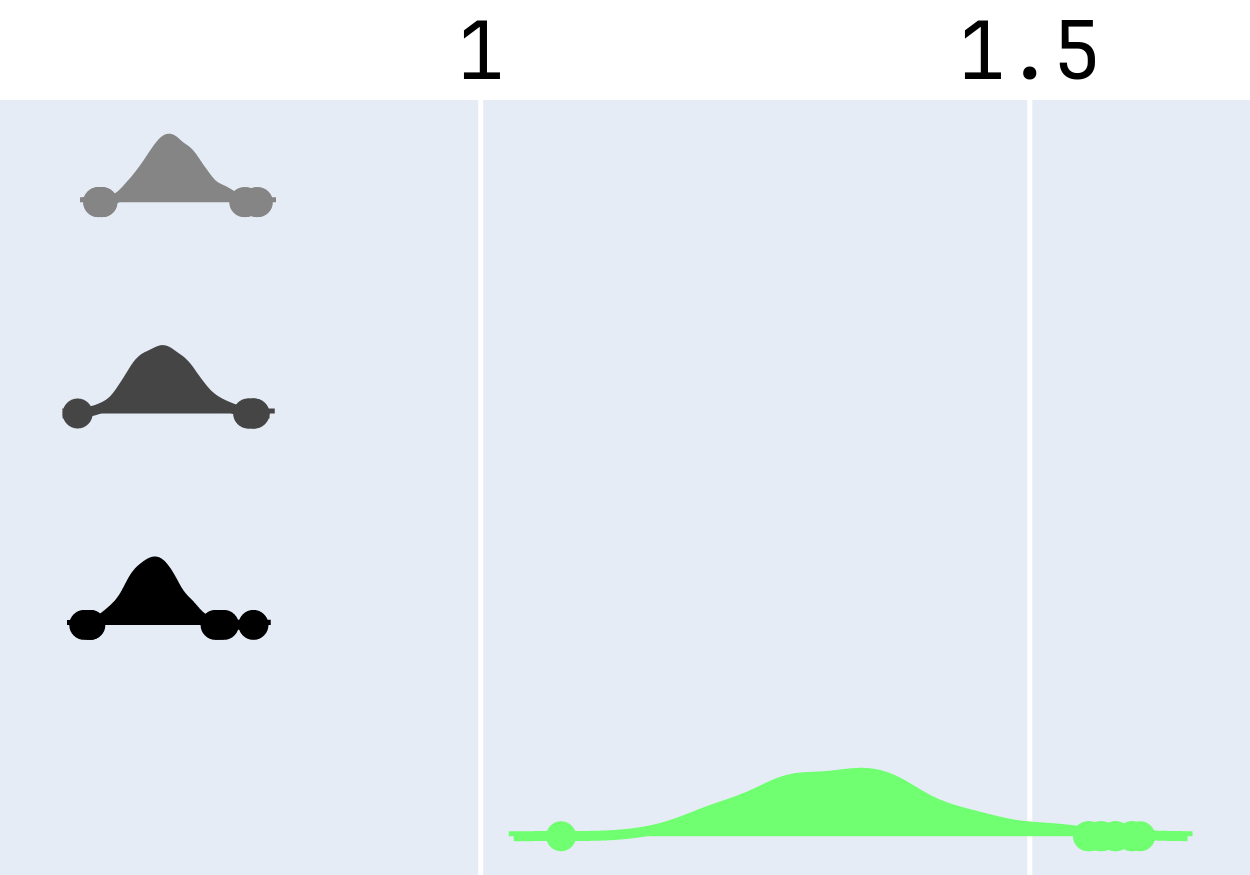}%
      \hspace{0.015\textwidth}%
      \includegraphics[width=0.31\textwidth]{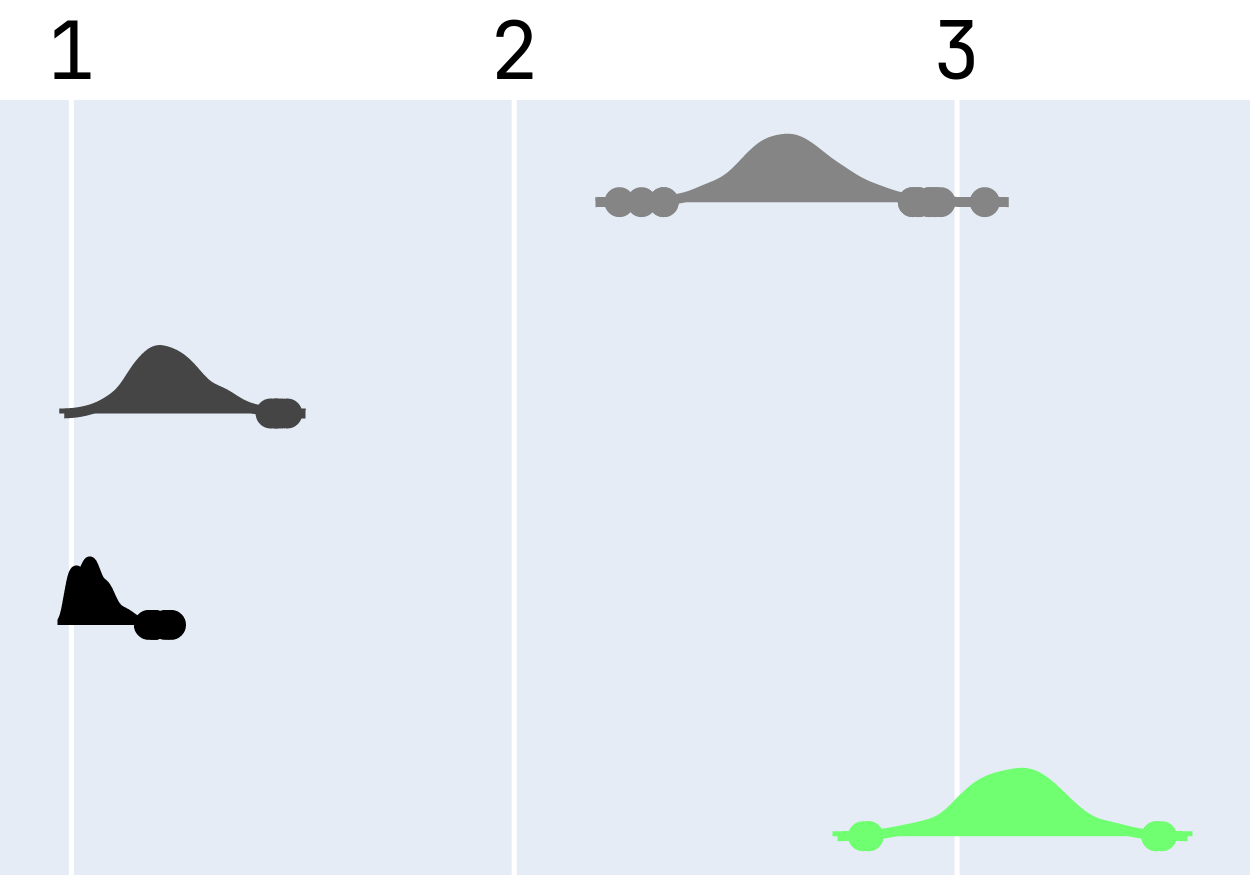}
    \end{minipage}
  \end{minipage}
  \label{fig:npsk_p3}
}

\caption{Distributions and ranks of the loss functions based on three different performance measures. From left to right, the performance measures are: MSS, P-Loss, and Likert. Higher values indicate better performance. Rank colors are \colorbox{rank1}{\textcolor{black}{\textbf{1}}} \colorbox{rank2}{\textcolor{white}{\textbf{2}}} \colorbox{rank3}{\textcolor{white}{\textbf{3}}} \colorbox{rank4}{\textcolor{white}{\textbf{4}}}.}
\label{fig:npsk_all}
\end{figure*}

In Figure~\ref{fig:npsk_all}, we use color-coded violin plots to visualize the best performers for each program. The corresponding colors for each rank are
\colorbox{rank1}{\textcolor{black}{\textbf{1}}} \colorbox{rank2}{\textcolor{white}{\textbf{2}}} \colorbox{rank3}{\textcolor{white}{\textbf{3}}} \colorbox{rank4}{\textcolor{black}{\textcolor{white}{\textbf{4}}}}.

\newcommand{\markdiff}[1]{\textbf{\makebox[0pt][r]{\textbf{*}}#1}}

\begin{table*}[p]
\centering
\caption{Ranks for each synthesis method (rows) under the three evaluation metrics (columns), across four targets (\BPNoise, \AddSineSaw, \AmpMod, \FMMod). Values marked with an asterisk are different from the Likert score hearing ranks.}
\small
\begin{tabular}{|c|ccc|ccc|ccc|ccc|}
\hline
\textbf{Function} 
  & \multicolumn{3}{c|}{\BPNoise}
  & \multicolumn{3}{c|}{\AddSineSaw}
  & \multicolumn{3}{c|}{\AmpMod}
  & \multicolumn{3}{c|}{\FMMod} \\
\cline{2-13}
  & MSS & P-LOSS & Hearing 
  & MSS & P-LOSS & Hearing 
  & MSS & P-LOSS & Hearing 
  & MSS & P-LOSS & Hearing \\
\hline

\textbf{SIMSE} 
  & 1    & \markdiff{2}   & 1    
  & 4    & 4    & 4    
  & \markdiff{3}   & 4    & 4    
  & 2    & 2    & 2    \\
\textbf{L1}    
  & 2    & \markdiff{1}   & 2    
  & 2    & 2    & 2    
  & 2    & 2    & 2    
  & 3    & 3    & 3    \\
\textbf{JTFS}  
  & 3    & \markdiff{4}   & 3    
  & 1    & 1    & 1    
  & \markdiff{4}   & 3    & 3    
  & 4    & 4    & 4    \\
\textbf{DTW}   
  & \markdiff{4}   & 3    & 3    
  & 3    & 3    & 3    
  & 1    & 1    & 1    
  & 1    & 1    & 1    \\

\hline
\end{tabular}
\label{tab:combined_ranks}
\end{table*}

\subsubsection{\BPNoise}
For this synthesizer program, the spectrogram-based models performed the best. This makes intuitive sense, as the visual effects of a band-pass filter on white noise are readily apparent in a spectrogram. As shown in Figure~\ref{fig:npsk_p0}, manual hearing test and MSS selected \SIMSESpec~as the best performer and \LoneSpec~as the second-best performer, while P-Loss gave the reverse order.

\subsubsection{\AddSineSaw}
As shown in Figure~\ref{fig:npsk_p1}, JTFS was the best performing loss function in all evaluation methods. Moreover, all evaluation methods produced identical results. 

\subsubsection{\AmpMod}
P-Loss and manual hearing tests yield identical rankings, and MSS results vary for ranks 2 to 4. It is not surprising that MSS results differ from the hearing tests, as we saw in Table~\ref{tab:kruskal_auto}, MSS showed no significant differences between groups. As shown in Figure~\ref{fig:npsk_p2}, all models selected \DTWEnv~as the best performer. This may be due to \DTWEnv's focus on periodic changes in loudness.

\subsubsection{\FMMod}
\DTWEnv~is again the best performer here. As shown in Figure~\ref{fig:npsk_p3}, all methods of analysis gave identical rankings to all programs.

\subsection{Consistency In Rankings}
\label{sec:consistency_in_rankings}
\subsubsection{Manual Ranks} Agreement between the manually assigned scores is measured using Spearman’s rank correlation, which provides both a correlation coefficient ($\rho$) ranging from –1 (perfect negative correlation) to 1 (perfect positive correlation), and a p-value testing the null hypothesis of no correlation~\cite{spearman1987proof,rebekic2015pearson}. Across all programs, the correlation was very strong ($\rho = 0.86$, $p < 10^{-180}$). Per-program correlations were also very strong: $\rho = 0.71$ for \BPNoise{} ($p < 10^{-25}$), $\rho = 0.64$ for \AddSineSaw{} ($p < 10^{-19}$), $\rho = 0.84$ for \AmpMod{} ($p < 10^{-43}$), and $\rho = 0.85$ for \FMMod{} ($p < 10^{-44}$). 

Table~\ref{tab:combined_ranks} shows the SNPK rankings of bootstrapped evaluation results for each program. Both MSS and P-Loss gave a different rank from the hearing results in 3 out of 16 cases, which shows consistency between automatic and manual hearing tests, at least for the simple programs used in this work. We also observe that top ranks were consistent across performance evaluation methodologies, with the exception of P-Loss in \BPNoise, which narrowly picks a different spectrogram-based loss function.

The experiment results provide answers regarding the main hypothesis of this work as well as the secondary questions. In the following sections, we will discuss our key findings, caveats, and make recommendations for future research.


\section{Discussion}
\subsection{Key Findings}
The most important takeaway regarding our main hypothesis is that the loss function that yields the best sound-matching outcomes varies depending on the synthesizer program. This is likely due to the interaction between how the parameters of the synthesizer influence the sound, and the core sonic features used by the loss function.

\textbf{Q1}: To what extent do automatic evaluation metrics agree with manual listening tests? 
\\We see somewhat consistent results between the rankings assigned by manual hearing tests (which we take as the ground truth), and automatic measures of P-Loss and MSS. With the exception of P-Loss in \BPNoise{} narrowly selecting a different spectrogram based loss, all measures of performance selected the same top performer. However, given the simplicity of our programs and the occasional deviations, we cannot conclude that these automatic measures are substitutes for human hearing tests. As many previous works have done (see Table~\ref{tab:summary}), the use of MSS and/or P-Loss as general loss functions does seem appropriate in general cases, however, the use of listening tests or custom made loss functions in specific contexts remains necessary. 

\textbf{Q2}: Are DTW and SIMSE effective measures of loss? If so, in what contexts?\\
\DTWEnv{} consistently outperformed other measures in amplitude-modulated synthesis, where envelope alignment is critical. \SIMSESpec{} performed best in subtractive synthesis, where scale-invariance captures noise-filtering effects more robustly than L1-based measures. These results suggest that the advantages of loss functions are context-dependent, underscoring the need for further exploration of creative differentiable similarity measures.

\textbf{Q3}: Can iterative differentiable optimization be applied effectively with synthesizers using classical DSP functions? \\
The approach of designing DSP functions in Faust and transpiling to differentiable Faust code was convenient, yielded meaningful outputs, and enabled comparative evaluation of losses. We were able to run our 1200 experiments within 72 hours on a laptop without a GPU\footnote{Lenovo ThinkPad T480, i5-8350U CPU at 1.70GHz, 32 GB RAM.}.

\subsection{Practical Recommendations}
    Our findings not only provide answers to the research questions proposed in the introduction, but also serve as guidelines for practitioners in selecting appropriate similarity measures and synthesis methods for future experiments.
\begin{itemize}
    \item In lieu of listening tests, MSS or P-Loss are generally useful for large-scale benchmarking. However, we encourage confirmation of such results with manual listening, particularly if specificity is important. 
    \item For amplitude-modulated synthesis, DTW-based losses such as \DTWEnv{} are recommended over spectrogram differences and JTFS.
    \item  For subtractive/noise-filtered synthesis, spectrogram based losses are most effective. Based on our results, SIMSE appears to be the better method of spectrogram comparisons relative to L1, but the amplitude agnosticism of SIMSE likely plays a role in its poor performance in additive synthesis.  
    \item Iterative differentiable optimization via the Faust-to-JAX pipeline is a viable strategy for defining various differentiable DSP functions, requiring only modest hardware resources and providing a more natural approach to synthesizer definition.
    \item In our amplitude modulation experiments, \JTFS{} did not outperform other measures, contradicting prior claims of its superiority for mesostructures~\cite{vahidi2023mesostructures}. Thus, our results call into question its effectiveness as a general mesostructure measure, and highlight a need for further research into its utility.
\end{itemize}

\subsection{Illustrating Loss Landscapes}
\label{sec:loss_landscape_examples}
To complement our quantitative analysis, we provide examples of loss landscapes that illustrate why specific losses succeed or fail in certain synthesis contexts. These are illustrative aids rather than conclusive evidence, and highlight the importance of \PeriodicLoss{} discussed earlier in Section~\ref{sec:lacking}.

We show the landscapes of the loss functions with simplified versions of \BPNoise{} (HP-Noise) and \AmpMod{} (S-Noise-AM), each reduced to a single parameter for one-dimensional visualization. HP-Noise uses only a high-pass filter with cut-off between 100–20,000 Hz, while S-Noise-AM varies only the modulation rate between 0.1–20 Hz. Loss values are normalized to 0–1 to allow comparison across functions.

\begin{figure*}[ht]
    \centering
    \begin{minipage}[t]{0.48\textwidth}
        \centering
        \includegraphics[width=\linewidth]{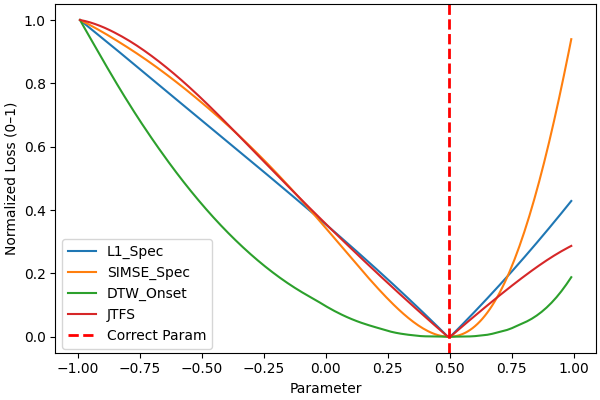}
        \caption{Loss landscapes for \BPNoise{} with only a high-pass filter parameter. 
        \LoneSpec{}, \SIMSESpec{}, and \JTFS{} show clear global minima near the correct parameter, while \DTWEnv{} remains flat around the target.}
        \label{fig:loss_landscape_noisebp}
    \end{minipage}%
    \hfill
    \begin{minipage}[t]{0.48\textwidth}
        \centering
        \includegraphics[width=\linewidth]{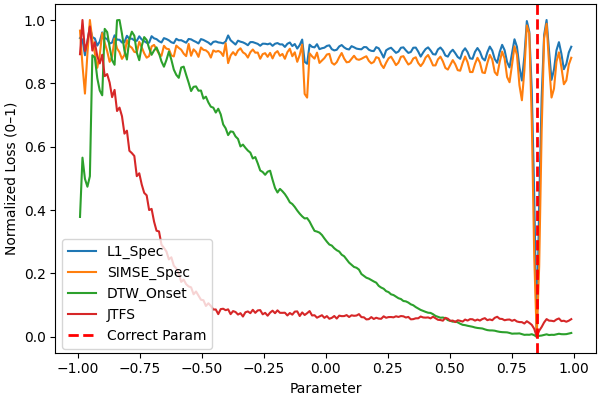}
        \caption{Loss landscapes for a simplified \AmpMod{} synthesizer with only amplitude modulation parameter. 
        \DTWEnv{} exhibits the smoothest and most informative landscape, explaining its superior performance in amplitude-modulated synthesis.}
        \label{fig:loss_landscape_ampmod}
    \end{minipage}
\end{figure*}

Figure~\ref{fig:loss_landscape_noisebp} shows the HP-Noise results. \LoneSpec{}, \SIMSESpec{}, and \JTFS{} exhibit clear minima at the correct parameter (red dashed line), whereas \DTWEnv{} remains relatively flat, making gradient-based optimization more difficult.

Figure~\ref{fig:loss_landscape_ampmod} shows the S-Noise-AM results. Here, \DTWEnv{} produces the smoothest and most informative landscape, while spectrogram losses are minimized near the target but lack consistent correlation with parameter distance. \JTFS{} remains largely flat. This aligns with \DTWEnv's superior performance in amplitude-modulated synthesis observed in our experiments.

\subsection{Applicability to More Complex Synthesizers}
The experiments in this study were conducted with differentiable synthesizers chosen to isolate fundamental synthesis principles, each with two parameters. While this level of simplicity facilitates controlled comparisons, it does not fully capture the richness of real-world synthesizers, which can have hundreds of parameters with various routes. As a result, the extent of the generalizability of our findings does require further research. 

Nonetheless, we believe that the insights here can extend naturally to more complex domains. Depending on the feature of interest, the observed dependence of loss performance on synthesis method is likely to remain stable even with added complexities such as layers of modulation, filtering, or nonlinearity. 

\DTWEnv{} measures very specific features of audio, and its use in the recommended settings---where low frequency amplitude modulation matching is required---would likely yield the desired results regardless of synthesizer complexity. Likewise, the use of \SIMSESpec{} or \LoneSpec{} for matching filter cut-offs is likely to be successful regardless of the underlying sound. 




\section{Summary, Weaknesses, and Conclusion}
\label{sec:summary_conclusion}
\textbf{Summary:} Sound-matching is an umbrella term for the algorithmic programming of audio synthesizers, often with the goal of assisting sound designers. Here we provided a history of sound-matching, major issues in the field, and discussed the importance of ``differentiable iterative sound-matching'' as a natural extension of current literature.

The main hypothesis tested here is whether the performance of differentiable loss functions is influenced by the synthesis techniques used for sound-matching. We conducted systematic iterative sound-matching experiments by combining four different loss functions with four different sound synthesis programs. We ranked the performance of the iterative sound-matching pipelines for every loss function and program, and observed that the success of the pipeline (that is, how closely the output sound matches the target sound) is program dependent. In other words, different synthesizer programs work best with different loss functions. Notably, we see that our novel use of DTW and SIMSE based differentiable loss functions (\DTWEnv{} and \SIMSESpec) can outperform what are regarded as the SOTA loss functions in 3 of 4 cases, although their success is highly synthesizer dependent. 

P-Loss and MSS have frequently been used as automatic performance measures, yet their ``preference'' has rarely been compared to human rankings. We observed that automatic performance measures and manual listening tests were generally in agreement, despite this, manual verification of sound-matching results remains a necessity in non-general experiments due to occasional divergences between automatic and manual tests.

\textbf{Weaknesses:} 
While we cannot prove that a universally best similarity measure does not exist, we can advocate for more creativity in the field. Compared to previous work, we presented a more cohesive approach to iterative sound-matching which utilizes a variety of loss functions and synthesis methods. However, there are many other methods of synthesis and sound-similarity that can be combined in practically infinite ways. Due to this large search space, we set arbitrary parameters for the various signal processing functions. We used bare-bones versions of STFT and JTFS with fixed parameters. We did not test complex synthesizers using parallel and sequential DSP functions. Arbitrary hyperparameters such as learning rate and max number of iterations were selected for the DL pipeline, and only the RMSProp optimizer was tested. 

 Like the majority of previous works, this work utilizes in-domain sounds; that is, the target sound is made by the synthesizer, and the parameters are already known. This simplifies the issues of measuring sound-similarity, but it is not a realistic scenario for practical sound-matching. This problem is left for future work, discussed in the next section.

\label{sec:future}
\textbf{Future Work: } The problem of periodic loss-landscapes, noted by many previous works, remains unaddressed~\cite{turian2020sorry,vahidi2023mesostructures,uzrad2024diffmoog,bruford2024synthesizer}. Perhaps this problem emerges due to the periodic nature of sound, which requires better loss landscape navigation methods. An optimizer that is more aware of fluctuations in the gradients would perhaps lead to better solutions than simple gradient descent. Viewing the sound-matching problem as ``the navigation of an agent from an arbitrary point in a gradient field to a target'' closely resembles many classical problems in the field of reinforcement learning (RL)~\cite{sutton2018reinforcement}. Naturally, the application of RL and other heuristic search techniques to the problem of iterative sound-matching would be an important contribution.

Contemporary works often involve the application of domain specific and computationally expensive loss functions~\cite{han2023perceptual,uzrad2024diffmoog}, use of large neural networks~\cite{hershey2017cnn,cramer2019look}, or complex ensemble methods~\cite{turian2022hear}. Such models are useful but intractable; furthermore, training them requires the definition of simpler loss functions, which emphasizes the need for further development of differentiable and expressive loss function implementations. 

Like nearly all previous work in sound-matching, the main measure of success here was the accurate \textit{replication} of sounds (or synthesizer parameters) rather than \textit{imitation} of sounds outside of the synthesizer's domain. Replication of arbitrary features of sounds is an important component of sound-design, though it is much harder to define or measure. Future works could explore loss functions which only measure certain characteristics of sound (such as \DTWEnv), and whether they can pave the way for better imitation in sound-matching.

\bibliography{references}

\end{document}